\documentclass[conference]{IEEEtran}
\IEEEoverridecommandlockouts

\usepackage{xcolor}
\usepackage{placeins}
 \usepackage{amsmath,amssymb,amsfonts}
 \usepackage{makecell}
\usepackage{subcaption}
\usepackage{multirow}
\usepackage{graphicx}
\usepackage[space]{grffile}
\usepackage{mathtools, nccmath}
\usepackage{dblfloatfix}
\usepackage{amsmath}
\usepackage{dsfont}
\usepackage{wrapfig}
\usepackage{algorithm}
\usepackage{float}
\usepackage{algpseudocode}
\usepackage{algcompatible}
\usepackage{mathtools}
\usepackage{algorithmicx}
\usepackage{pifont}%
\usepackage{makecell}
\newcommand{\RomanNumeralCaps}[1]
    {\MakeUppercase{\romannumeral #1}}

\def\BibTeX{{\rm B\kern-.05em{\sc i\kern-.025em b}\kern-.08em
    T\kern-.1667em\lower.7ex\hbox{E}\kern-.125emX}}
\begin{document}

\title{A Data-Driven Frequency Scaling Approach for Deadline-aware Energy Efficient  Scheduling on  Graphics Processing Units (GPUs)}

\author{\IEEEauthorblockN{Shashikant Ilager, Rajeev Muralidhar, Kotagiri Rammohanrao, Rajkumar Buyya}
\IEEEauthorblockA{\textit{Cloud Computing and Distributed Systems (CLOUDS) Laboratory} \\
\textit{School of Computing and Information Systems, 
The University of Melbourne, Australia}\\
}
}

\maketitle

\begin{abstract}
Modern computing paradigms, such as cloud computing, are increasingly adopting GPUs to boost their computing capabilities primarily due to the heterogeneous nature of AI/ML/deep learning workloads. However, the energy consumption of GPUs is a critical problem. Dynamic Voltage Frequency Scaling (DVFS) is a widely used technique to reduce the dynamic power of GPUs. Yet, configuring the optimal clock frequency for essential performance requirements is a non-trivial task due to the complex nonlinear relationship between the application's runtime performance characteristics, energy, and execution time. It becomes more challenging when different applications behave distinctively with similar clock settings. Simple analytical solutions and standard GPU frequency scaling heuristics fail to capture these intricacies and scale the frequencies appropriately. In this regard, we propose a data-driven frequency scaling technique by predicting the power and execution time of a given application over different clock settings. We collect the data from application profiling and train the models to predict the outcome accurately. The proposed solution is generic and can be easily extended to different kinds of workloads and GPU architectures. Furthermore, using this frequency scaling by prediction models, we present a deadline-aware application scheduling algorithm to reduce energy consumption while simultaneously meeting their deadlines. We conduct real extensive experiments on  NVIDIA GPUs using several benchmark applications. The experiment results have shown that our prediction models have high accuracy with the average RMSE values of 0.38 and 0.05 for energy and time prediction, respectively. Also, the scheduling algorithm consumes $15.07\%$ less energy as compared to the baseline policies. 
\end{abstract}


\section{Introduction}

Graphics Processing Units (GPUs) have become ubiquitous in modern computing paradigms and platforms, such as Cloud computing and supercomputing environments, due to their massive computational capabilities. Furthermore, the  Single Instruction Multiple Data (SIMD)  architecture of GPUs is ideally suitable for many parallel and compute-intensive scientific and business workloads \cite{luebke2006gpgpu,jeon2018multi}. These advantages manifested into the deployment of a large number of  GPU clusters in many data centers, including  Top500  supercomputers and also in the public Clouds \cite{ top500, keckler2011gpus, exposito2013general, ilager2019gpu}. In spite of this increased usage, the power consumption of GPUs has become a significant bottleneck for designing hyper-scale GPU systems  \cite{bridges2016understanding, ebrahimi2014review}. On the other hand, GPU  workloads are more sensitive to their Quality of Service (QoS) constraints requiring faster execution and thus spending more energy. Therefore,  energy-efficient workload management with QoS satisfaction is exceedingly essential.

 Dynamic Voltage Frequency Scaling (DVFS) is a popular technique to reduce active power by varying the GPU frequencies \cite{mei2013measurement, ge2013effects, guerreiro2019modeling}. The modern  GPUs have two frequency domains, core and memory, each with many numbers of frequency ranges. While former regulates the  Streaming Multiprocessors (SM) (including register, texture cache, shared memory, and l2 cache), and the latter governs bandwidth of DRAM \cite{guerreiro2019modeling}. For instance,  NVIDIA Tesla P100 GPU  supports one memory frequency (715 MHz)  and 62 core frequencies ([544-1328] MHz), and NVIDIA GTX 980 supports four memory frequencies ([3505-324] MHz) and  87  different core frequencies( [135-1428] MHz ) with the total number of  267 possible frequency combinations. A  particular combination of memory and core frequency can be set using the NVIDIA Management Library (NVML). However, the principle DVFS notion- higher frequency range increases the performance requiring more power, while lower frequency consumes less power by decreasing the performance do not hold in all the scenarios \cite{guerreiro2019dvfs}.
In addition,  different GPU application kernels behave differently concerning energy and performance with the frequency settings due to their different resource footprints and the intricate instruction execution patterns. Thus, due to such non-linear dependencies, estimating and optimally scaling the frequencies for a given application is non-trivial. 

 Furthermore, frequency scaling becomes more challenging when a scheduler needs to schedule multiple applications with their deadline requirements. In such a case, the scheduler should not only identify the energy-efficient frequency combinations, but it also needs to take care of the application's execution time. Such scenarios are highly prevalent in real-time HPC and cloud environments \cite{lee2011heterogeneity, capodieci2018deadline}. 
 However, existing analytical and heuristic-based GPU frequency scaling \cite{Wang2019, Fan2019} methods are inefficient as they fail to capture the complex non-linearity between the frequency settings, performance, and power. To that end, data-driven DVFS scaling is a promising technique that is built using actual measurements. Models built using such methods can accurately scale the frequencies based on application demands\cite{Bridges2016}. Moreover, once the model is trained, the new applications can be scheduled on-the-fly with minimum profiling data.

 In this paper, we present a data-driven approach for frequency scaling by observing key architectural, power, and performance counters and predicting the estimated application power and execution time. In addition, guided by these prediction models, we propose a deadline-aware energy-efficient scheduling algorithm that accurately scales the GPU frequency according to the application requirements. We use twelve applications for evaluation from two standard GPU benchmarking suites, Rodinia \cite{rodinia_che} and Polybench \cite{polybench_gray}. The training data is generated from profiling the applications using  \textit{nvprof}, a standard profiling tool from NVIDIA. Furthermore,  several machine learning models are explored to accurately predict the energy and execution time of applications for the given frequency domains. Based on the experimental results,  CatBoost, an ensemble-based gradient boosting learning model, is chosen for prediction modeling. We implement the prototype scheduling system and evaluate the proposed techniques on real platforms. The experimental results conducted on the NVIDIA GPU device, Tesla P100 (Pascal micro-architecture), have shown that our prediction models have high accuracy and  proposed scheduling algorithm consumes less energy as compared to the baseline algorithms.

In summary, we make the following key \textbf{contributions}:
\begin{itemize}
    \item We propose a data-driven prediction model to accurately predict the energy and execution time of applications to assist the efficient frequency scaling configuration by observing key architectural, power, and performance counters and metrics.
    \item We design and present a deadline-aware energy-efficient application scheduling algorithm using the prediction models. 
    \item We implement a prototype system and evaluate the proposed solution on a real platform using standard benchmarking applications
    \item We show the efficiency and efficacy of our proposed solution with extensive experiments, and results are compared and analyzed with the existing state-of-the-art solutions. 
\end{itemize}

The remainder of the paper is organized as follows. Section \RomanNumeralCaps{2} describes the DVFS background, motivation of the work, and system model. Section \RomanNumeralCaps{3} presents the data-driven frequency scaling techniques. Section \RomanNumeralCaps{4} shows our proposed deadline aware energy-efficient scheduling algorithm. Section \RomanNumeralCaps{5} describes the performance evaluation with the analysis of the results. The related work is explained in Section \RomanNumeralCaps{6}. Finally, Section \RomanNumeralCaps{7}  draws the conclusions and future work.

\begin{figure}
  \captionsetup{justification=centering}
\begin{subfigure}[t]{0.24\textwidth}
\includegraphics[width=\linewidth]{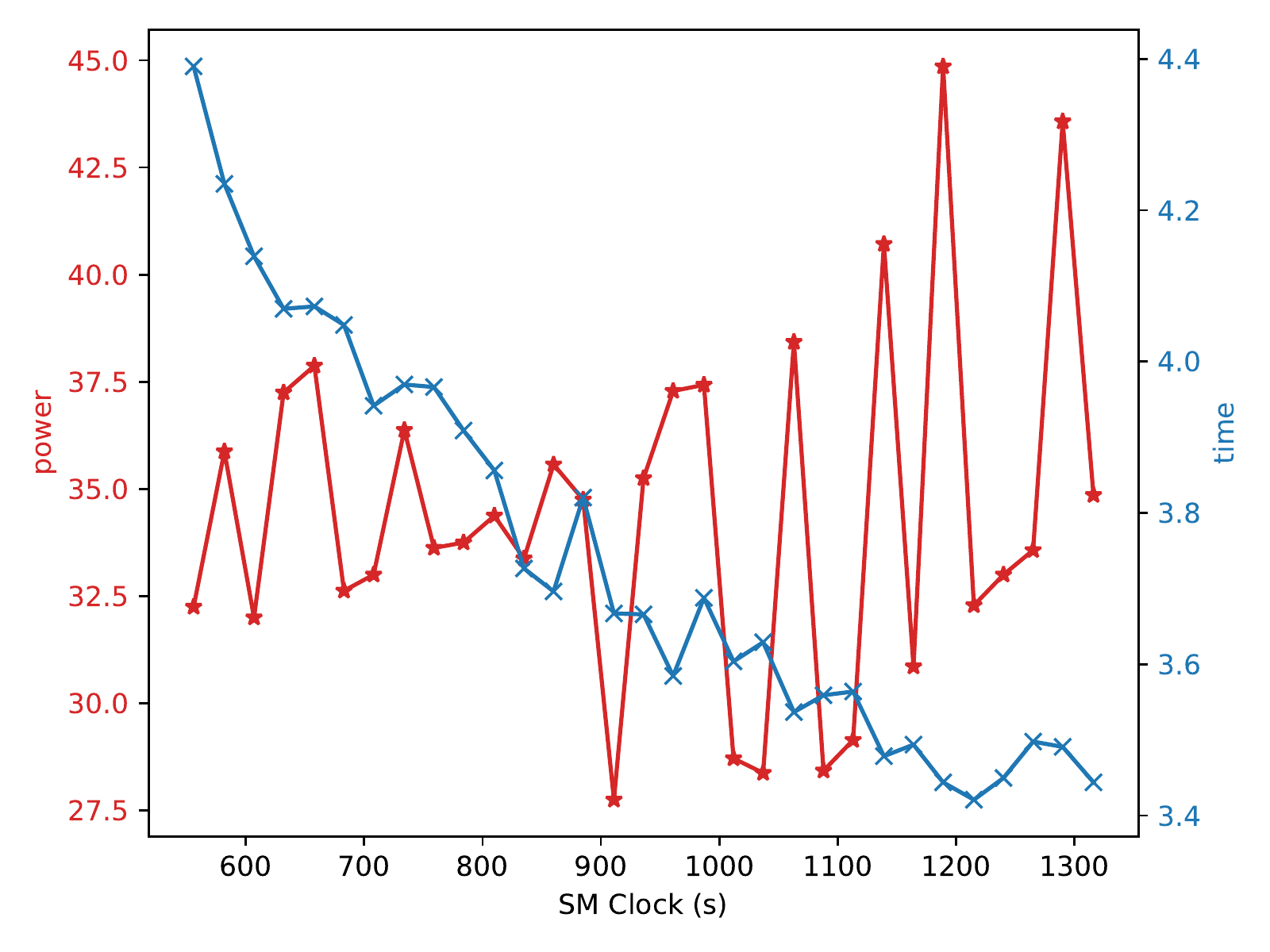} %
   \caption{lavaMD}
   \label{fig:motivation-lavaMD}
\end{subfigure}
\begin{subfigure}[t]{0.24\textwidth}
\includegraphics[width=\linewidth]{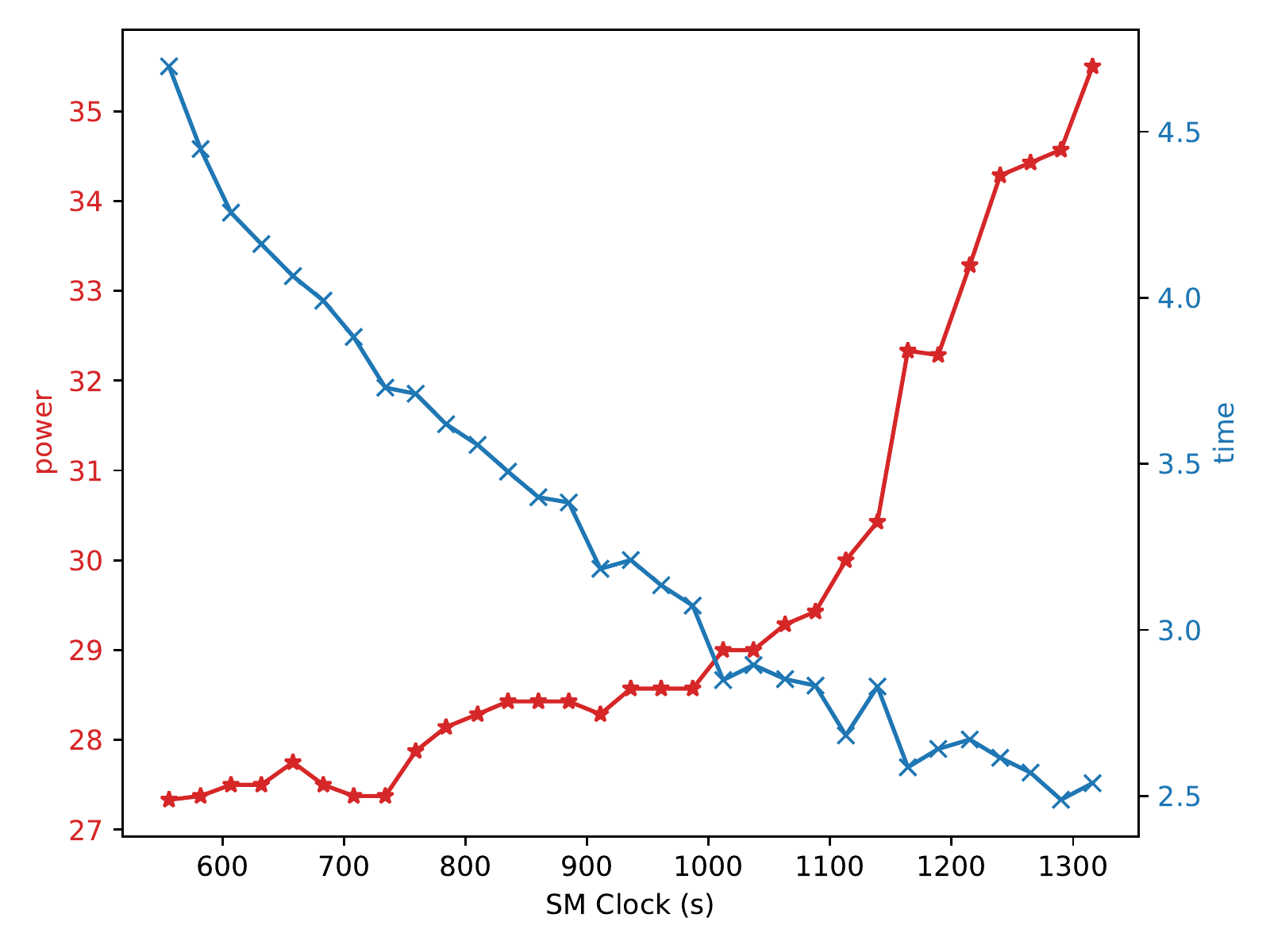}
   \caption{CORR}
   \label{fig:motivation-CORR}
\end{subfigure}
\begin{subfigure}[t]{0.24\textwidth}
\includegraphics[width=\linewidth]{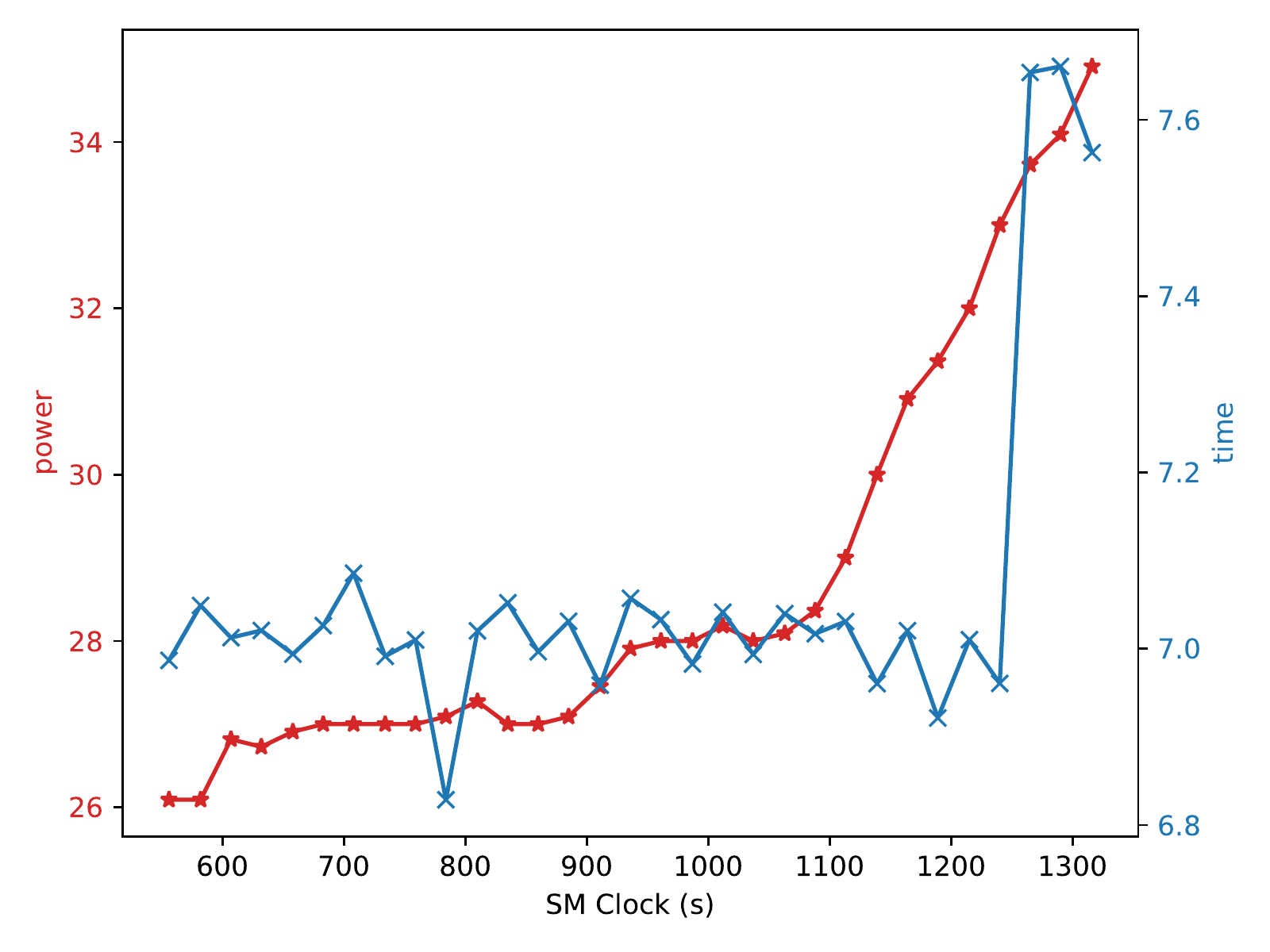}
   \caption{myocyte}
   \label{fig:motivation-myocyte}
\end{subfigure}
\begin{subfigure}[t]{0.24\textwidth}
\includegraphics[width=\linewidth]{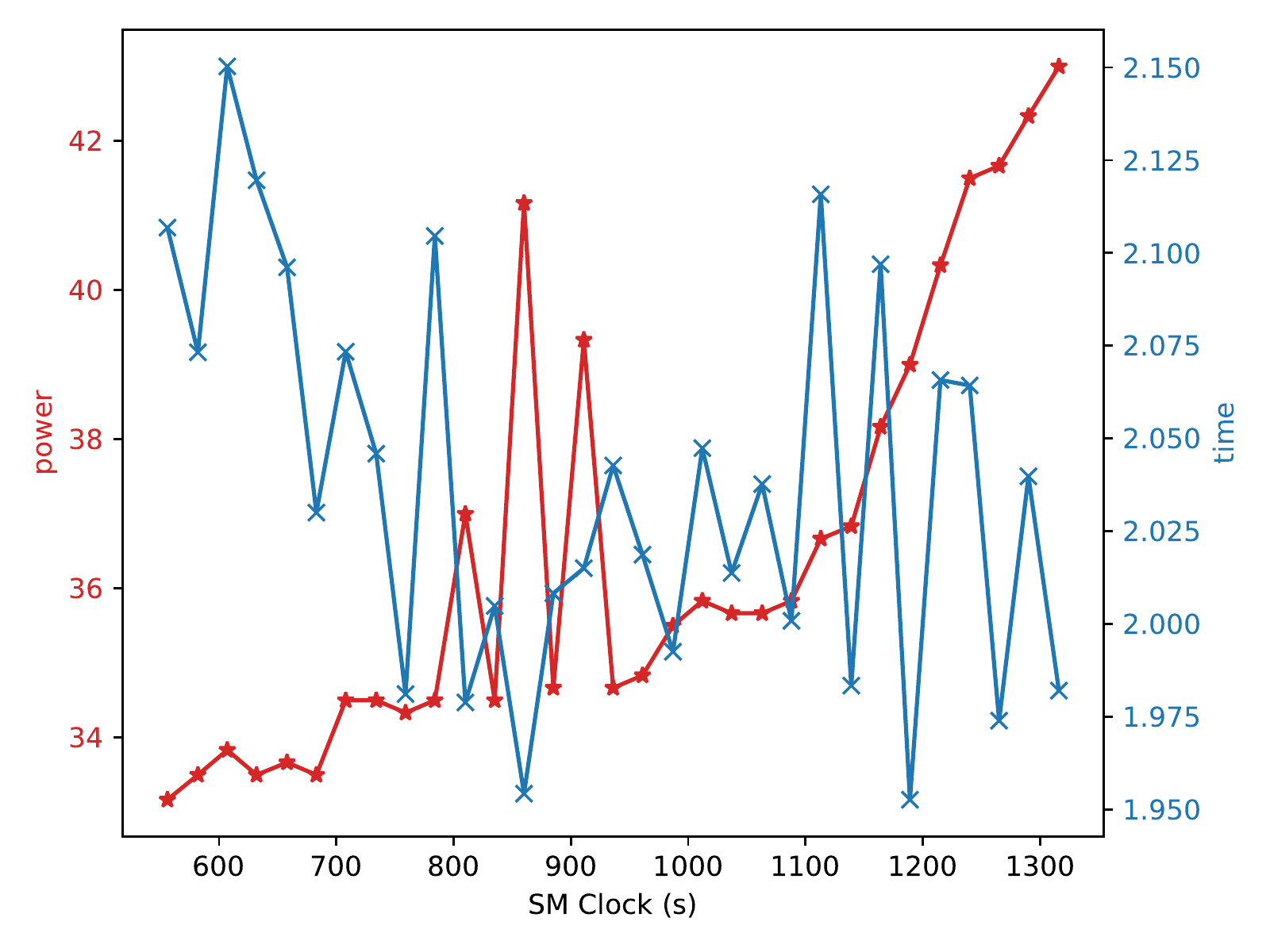}
   \caption{SYR2K}
   \label{fig:motivation-syr2k}
\end{subfigure}
\caption{Power, time and clock relationship of different applications}
   \label{fig:motivation}
\end{figure}

\section{Background Motivation and System Model}

 \subsection{GPU DVFS}
The power consumption of a GPU (or any semiconductor logic block in general) is a combination of dynamic  ($P_{dynamic}$) and static power ($P_{static}$).
The static power ($P_{static}$) is related to the leakage and energy consumed when the system is idle, and usually it is managed by hardware and/or software using different sleep states \cite{basu2018interference}. However, large amounts of energy is spent on the dynamic power ($P_{dynamic}$) which is proportional to the  run time of the workload. Performance management of GPU typically rely on the DVFS-based heuristics to regulate the dynamic power to save the energy. The frequency is normally regulated based  on the application's utilisation parameters or system's temperature threshold throttling mechanisms \cite{Bridges2016}. Thus, the dynamic power can defined as below:
 \begin{equation}
  \label{eq:dynpowerpower}
        \begin{aligned}
            P_{dynamic} \hspace{.2cm} \alpha  \hspace{.2 cm} V^{2}F
        \end{aligned}
\end{equation}
In Equation \ref{eq:dynpowerpower}, $F$ denotes the operating frequency, while $V$ denotes the supply voltage. Based on the current operating frequency, a combination of hardware and software changes the frequency (and thereby the underlying voltage); certain frequency ranges can share the same voltage level. Furthermore, GPUs have multiple frequency domains $F = \{f_{smclock}, f_{memclock}\}$, regulating hardware components related to streaming multiprocessor or graphics processor  and the memory components, respectively \cite{guerreiro2019dvfs}. Thus, considering that, usually, hardware logic manages the voltage based on operating frequency, we focus on benefiting from regulating frequency and scaling it based on application behaviors.
\subsection{Motivation}

Estimating the optimal frequency is a non-trivial problem due to the complex behaviors of applications regarding their energy consumption and execution time. To analyze this complexity, we plot the behavior of different applications towards energy and execution time by changing the core frequency of the GPU, as shown in Figure \ref{fig:motivation}. These executions are from NVIDIA Tesla P100 GPU that has one memory frequency and 62   core frequencies. So we only vary the core frequencies. In Figure \ref{fig:motivation}, we can observe that when the core frequency increases, energy consumption is not always linear. And also, the lavaMD (Figure \ref{fig:motivation-lavaMD}) application has a completely inconsistent response to frequency variations. Furthermore, some application produces different functionalities between certain frequency range, for application CORR (Figure \ref{fig:motivation-CORR}), energy consumption has a non-convex curve between [730-920] MHz.  Similar nonlinear behavior is present in Figure \ref{fig:motivation-myocyte} and \ref{fig:motivation-syr2k}, where execution time and energy consumption have some unexpected spikes and dips in their energy response. Moreover, optimising such non-linear non-convex  functions is an NP-hard problem \cite{allen2016variance}. Therefore, it is extremely challenging to find energy-efficient frequency combinations under application's QoS constraints. Simple analytical models that linearly regulate the core frequencies are inaccurate and inefficient to reduce the energy or increase the performance \cite{Bridges2016}. Motivated by these factors, we model the frequency scaling problem based on the data-driven methods.

\begin{table*}[t]
\centering
\begin{tabular}{llll}
\hline
\textbf{Application}  & \textbf{Domain/Description}                                                          & \textbf{Suite} & \textbf{Input}                        \\
\hline
particlefilter\_naive & Medical Imaging                                                                      & Rodinia        & -x 128 -y 128 -z 10 -np 1000          \\
particlefilter\_float & Medical Imaging                                                                      & Rodinia        & -x 128 -y 128 -z 10 -np 1000          \\
myocyte               & Biological Simulation                                                                & Rodinia        & 10000, 1000, 1                        \\
lavaMD                & Molecular Dynamics                                                                   & Rodinia        & -boxes1d 50                           \\
Backprop              & Pattern Recognition                                                                  & Rodinia        & 983040                                \\
SYRK                  & Symmetric rank−k operations                                                          & Polybench      & M 1024, N 1024                        \\
SYR2K                 & Symmetric rank−2k operations                                                         & Polybench      & M 2048, N 2048                        \\
GEMM                  & \begin{tabular}[c]{@{}l@{}}Matrix Multiply C = αA x B +\\ βC\end{tabular}            & Polybench      & NI 2048, NJ 2048, NK 2048             \\
COVAR                 & Covariance Computation                                                               & Polybench      & M 2048, N 2048                        \\
CORR                  & Correlation Computation                                                              & Polybench      & M 2048, N 2048                        \\
ATAX                  & \begin{tabular}[c]{@{}l@{}}Matrix Transpose and Vector\\ Multiplication\end{tabular} & Polybench      & NX 16384, NY 16384                    \\
2MM                   & 2 Matrix Multiplications (D=A.B; E=C.D)                                              & Polybench      & NI  4096, NJ  4096, NK  4096, NL 4096 \\
\hline
\end{tabular}
\caption{Description of applications}
\label{table:benchmark}
\end{table*}
\subsection{System Model}
\begin{figure}
  \captionsetup{justification=centering}
\includegraphics[width=\linewidth]{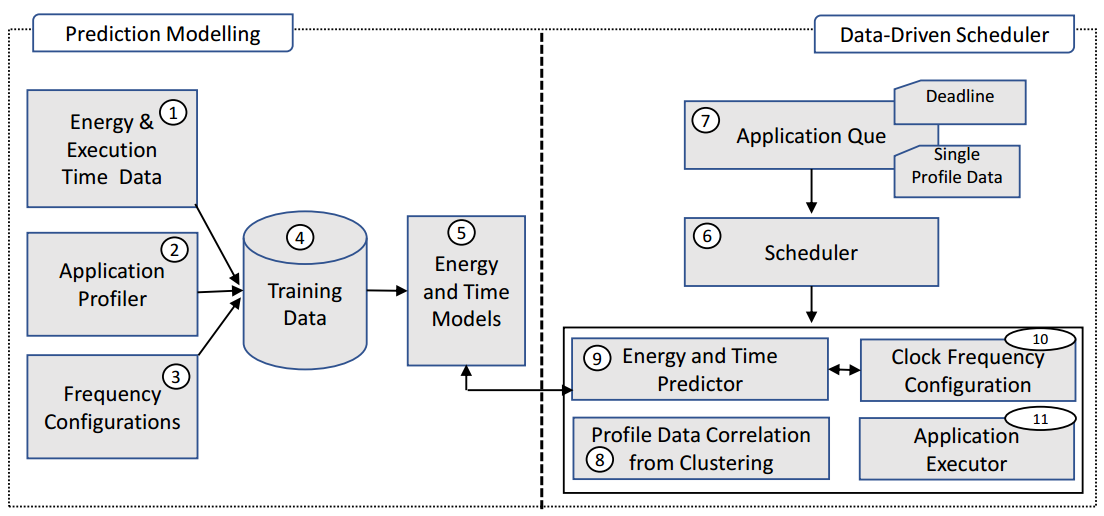} 
   \caption{System Model}
   \label{fig:system_model}
\end{figure}
A high-level system model of the proposed framework is shown in Figure \ref{fig:system_model}. It is broadly classified into two parts, \textit{predictive modeling} and \textit{data-driven scheduler}. In the first part, we collect the training data that consists of three parts, profiling information, energy-time measurements, and respective frequency configurations.  We then predict two entities for a given application and frequency configuration, i.e., energy consumption and execution time.  Afterward, in the second part, the new applications arrive with the deadline requirements and minimal profiling data from a default clock frequency execution. The scheduler finds correlated application data using the clustering technique, and this data is used for predicting the energy and execution time over all frequencies. Finally, based on the deadline requirements and energy efficiency, the scheduler scales the frequencies and executes the applications. The important components of this framework are explained in the following sections. 
 
\section{Data-Driven Frequency Scaling for GPUs}
In this section, we discuss the prediction models in detail, which include data collection, training, and model evaluation.

\subsection{Feature Collections}

\begin{table}[]
\begin{tabular}{ll}
\hline
\textbf{Power}                      & \textbf{Time}                  \\ \hline
sm                                  & sm                             \\
sm\_clock                           & l2\_tex\_read\_hit\_rate       \\
l2\_tex\_read\_hit\_rate            & l2\_tex\_read\_transactions    \\
tex\_cache\_throughput              & tex\_cache\_transactions       \\
ipc                                 & dram\_write\_transactions      \\
flop\_dp\_efficiency                & ipc                            \\
shared\_load\_throughput            & inst\_executed\_shared\_loads  \\
stall\_exec\_dependency             & gst\_efficiency                \\
stall\_inst\_fetch                  & inst\_replay\_overhead         \\
eligible\_warps\_per\_cycle         & inst\_executed\_shared\_stores \\
stall\_constant\_memory\_dependency & l2\_read\_throughput           \\
pcie\_total\_data\_transmitted      & gst\_throughput                \\
dram\_read\_transactions            & warp\_execution\_efficiency    \\
dram\_read\_bytes                   & dram\_read\_bytes              \\
issue\_slots                        & local\_store\_throughput       \\
l2\_tex\_write\_throughput          & gld\_efficiency                \\
inst\_bit\_convert                  & global\_store\_requests        \\
l2\_global\_load\_bytes             & stall\_memory\_throttle        \\
gld\_requested\_throughput          & dram\_utilisation              \\
pcie\_total\_data\_received         & inst\_fp\_32                   \\

\hline
\end{tabular}
\caption{Features used in energy and time prediction (top 20)}
\label{table:featuretable}
\end{table}
ML-based models are trained using real-time measurement data from the environment, and these models are used in the run time to predict the outcomes. In case of GPU energy and performance prediction, several existing studies rely either on static source code metrics \cite{Fan2019, Chau2017}, or on run time traces and profiling data \cite{wu2015gpgpu}. The profiling based method is most suitable due to its ability to gather the real resource footprints and hardware counters of applications on particular GPU architectures, which is crucial to estimate energy and time accurately.

The input to  our training model consists of (1) applications profiled features  $F = \{f1,f2,f3... fn\}$,  (2) respective GPU frequency pair  $\{ f_{smclock}, f_{memclock}\}$, and (3) energy and execution time measurements. The profiling features contain the information of an application's run time metrics related to its different hardware components utilization values, instruction counts, communication, and cache metrics, etc.

\textbf{Applications:} We have considered twelve different applications from two heterogeneous computing benchmark suites, Polybench \cite{polybench_gray} and Rodinia \cite{rodinia_che}. These two benchmark suites cover a wide range of applications. The Polybench suit covers many linear algebraic applications while the Rodinia suit covers different scientific applications. The details of these applications, including their domain and input specifications, are shown in Table \ref{table:benchmark}.

\textbf{Profiling:}  For profiling the applications, we use \textit{nvprof}, a standard profiling tool by NVIDIA for CUDA applications. Although NVIDIA has recently released new  \textit{nsigh-systems} tool kits for monitoring and profiling, they do not support many existing GPU architectures and CUDA versions, so we use \textit{nvprof}. We have also used the  \textit{nvidia-smi} toolkit, which is built on top of \textit{nvml} library, a C-based API for monitoring and managing various states of the NVIDIA GPU devices. This tool allows application users to set the supported GPU application clocks and also to measure the energy consumption metrics.

We have developed Python scripts to collect the profiling metrics that runs all the applications iteratively on different frequency domains supported by GPU.
Initially, we collect all the available metrics provided by \textit{nvprof} using \textit{--metrics all} argument and export the output in \textit{csv} format. The energy and execution time are gathered by running applications separately to avoid the effect of profiling on these metrics. 

We collect metrics from every alternative clock pair of the  Tesla P100 GPU from its supporting  62  combinations of core and memory frequencies to reduce the data collection time. It is important to note that,  some applications take up to ten minutes for each profiling session as  \textit{nvprof}  replays the application kernels over several passes to collect the metrics.  The \textit{nvprof}  provides more than 120 metrics of GPU counters for each execution. For the sake of brevity, we list the top twenty features that dominate in energy and execution time prediction in Table \ref{table:featuretable}. The details about selecting these features are explained in feature analysis Section \ref{Sec:FA}. Here, the features \textit{sm} (SM's utilisation level)  is collected from \textit{nvidia-smi dmon} API and remaining all are from \textit{nvprof}. 

These collected application profiling metrics, along with the frequency configuration, energy, and time metrics, constitute the total training data, which are then used to build predictive models.
\subsection{Prediction Models}

 When building any predictive models, it is often required to test the suitable candidate algorithms and adopt the model that works best for the given input training data and the problem domain. In this regard,  the prediction of energy and execution time requires regression-based machine learning (ML) models. We have investigated several suitable candidate ML algorithms, including  Linear Regression (LR), lasso-linear regression (Lasso), and Support Vector Regression (SVR). Also, we  explored ensemble-based gradient boosting techniques,  eXtreme Gradient Boosting (XGBoost), and CatBoost. The goal is to build energy and execution time prediction models for each GPU device to assist the frequency configuration. To that end,  prediction models are trained for two outputs, i.e., energy ($E$) and execution time($T$). The profiling data from all the applications are partitioned into training and testing datasets with 70\% and 30\%, respectively. 

The input feature set now includes a set of  tuples with each tuple having profiled features and frequency combination i.e,  $F = \{f_1,f_2,f_3, ... f_n\} \cup \{ f_{smclock}, f_{memclock}\}$ while the two models output predicted  energy consumption $P$ and execution time $T$. 

We use the sci-kit learn package [22] to implement the LR, Lasso, and SVR. For XGBoost and CatBoost, the standard python packages are used that are publicly available \footnote{https://catboost.ai/} \footnote{https://github.com/dmlc/xgboost}. A few of the profiling parameters from $nvprof$ are categorical, representing different components utilization as $low, mid$, and $high$. Among a total of 120 features collected, 15 features were categorical including $dram\_utilisation$, $double\_precision\_fu\_utilisation$.  Only numerical features are used in the models except for CatBoost. However, CatBoost is specifically designed to work with the categorical or mixed data, and it has an efficient way of representing the categorical variables. Here, the categorical features are transformed into numerical features by the technique of order target statistics. 

The parameters for each of the algorithms are the default and self-explanatory in our implementation. To avoid over-fitting of the models, we adopt the leave-one-out cross-validation, where we exclude the data from a  particular application in training and evaluate this model with the excluded application's test data. This helps to assess the robustness of models, and proven efficiency will help to use these offline trained models  for new applications without retraining the models.

The goodness of fit is measured using the Root Mean Square Error (RMSE) metric, which is a standard evaluation metric in regression-based problems [23]. The RMSE is defined as: 
\begin{equation}
  \label{eq:rmse}
        \begin{aligned}
        RMSE = \sqrt{\frac{1}{n} \sum_{i=0}^{n}\left ( y_i - y^i \right )^2 }
        \end{aligned}
\end{equation}

In Equation \ref{eq:rmse}, $y_i$ is the observed value, $y^i$ is the predicted output variable by prediction model, and $n$ is the total number of predictions. The value of RMSE represents the standard deviation of the residuals or prediction errors. It also indicates how far are the data points from the model fitted line. Thus, lower RMSE values are preferred. 

\begin{figure}[t]
  \captionsetup{justification=centering}
\begin{subfigure}[t]{0.47\textwidth}
\includegraphics[width=\linewidth]{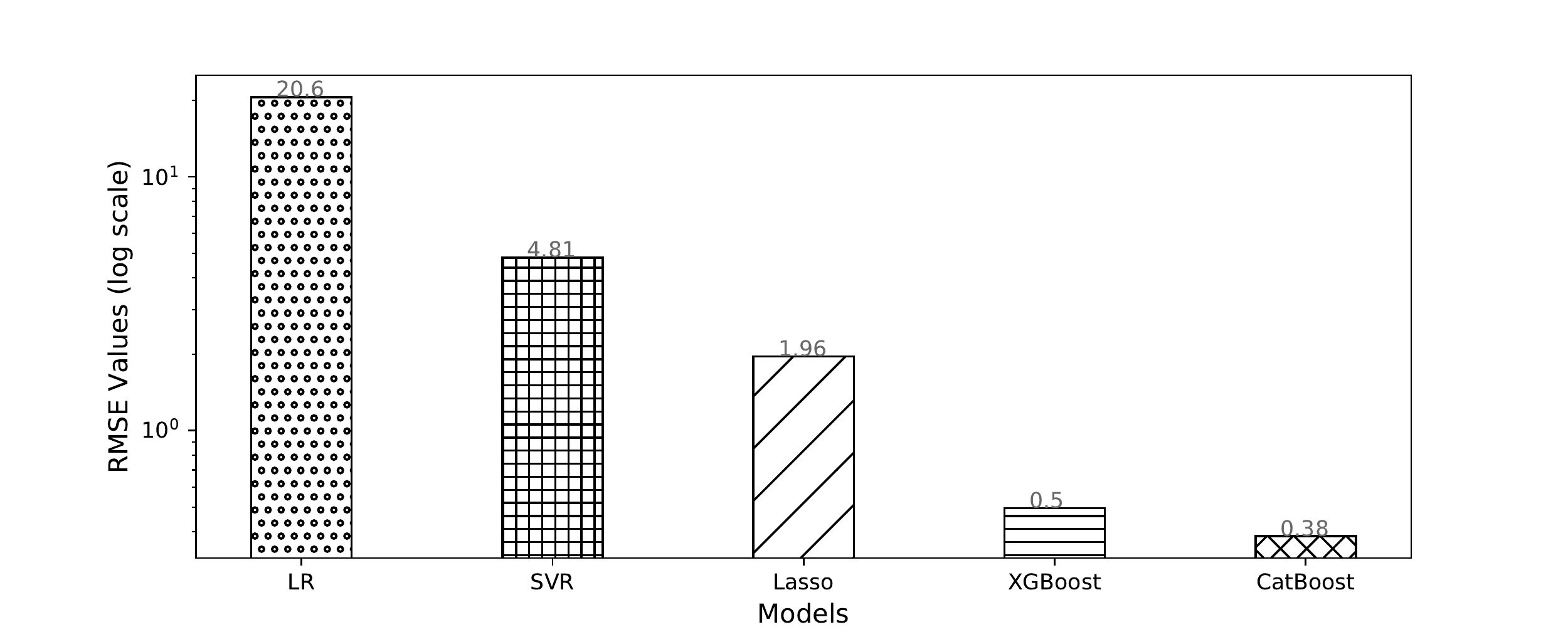} 
   \caption{Energy prediction }
   \label{fig:rmseenergy}
\end{subfigure}
\begin{subfigure}[t]{0.47\textwidth}
\includegraphics[width=\linewidth]{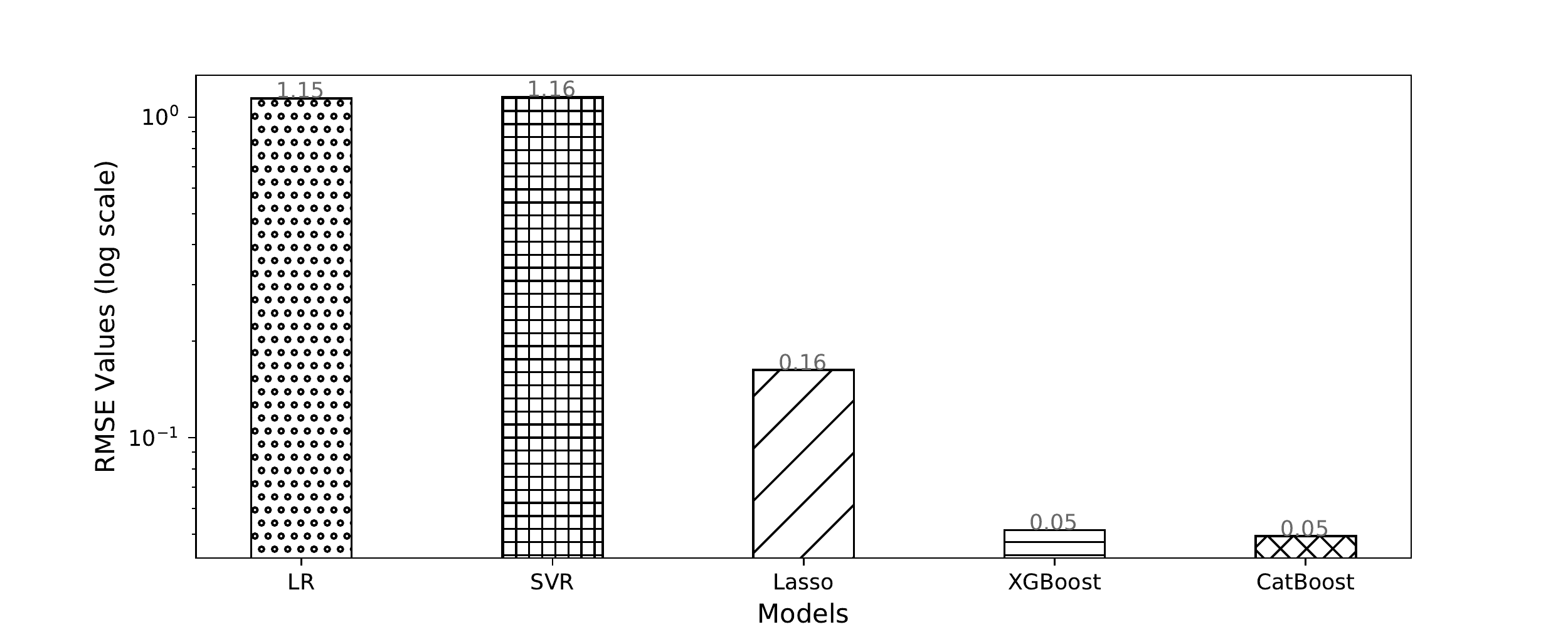}
  \caption{Time prediction}
   \label{fig:rmsetime}
\end{subfigure}
   \caption{ Performance of different models for energy and execution time prediction (lower RMSE value is preferred)}
   \label{fig:rmse}
\end{figure}

The performance of different algorithms is shown in  Figure \ref{fig:rmse}. Here, we can observe that the CatBoost has the lowest RMSE value of 0.38 in energy prediction, indicating residuals or prediction errors are less, and its predictions are more accurate. We observed that linear models like LR, SVR, and Lasso perform worst in estimating energy and slightly better in predicting the execution time. It is because energy consumption has more non-linearity with the input features than the execution time, and simple linear models do not perform well at it. While in execution time prediction (Figure \ref{fig:rmsetime}), both XGBoost and CatBoost has equal performance (RMSE value of 0.05). As the performance of the CatBoost is promising in both models, we choose it as our prediction algorithm.

 We perform hyperparameter tuning to further optimize the CatBoost model; we use the grid search technique over the parameter space. The results of the grid search are shown in Table \ref{table:girdsearch}. Here, the parameters $iterations$ and $depth$  decides the number and size of the decision trees while $learning\_rate$  is used for reducing the gradient step. The $l2\_leaf\_reg$ represents the coefficient at the L2 regularisation term of the cost function.

\subsection{Feature Analysis}\label{Sec:FA}

\begin{figure*}
  \captionsetup{justification=centering}
  \begin{subfigure}[t]{0.48\textwidth}
\includegraphics[width=\linewidth]{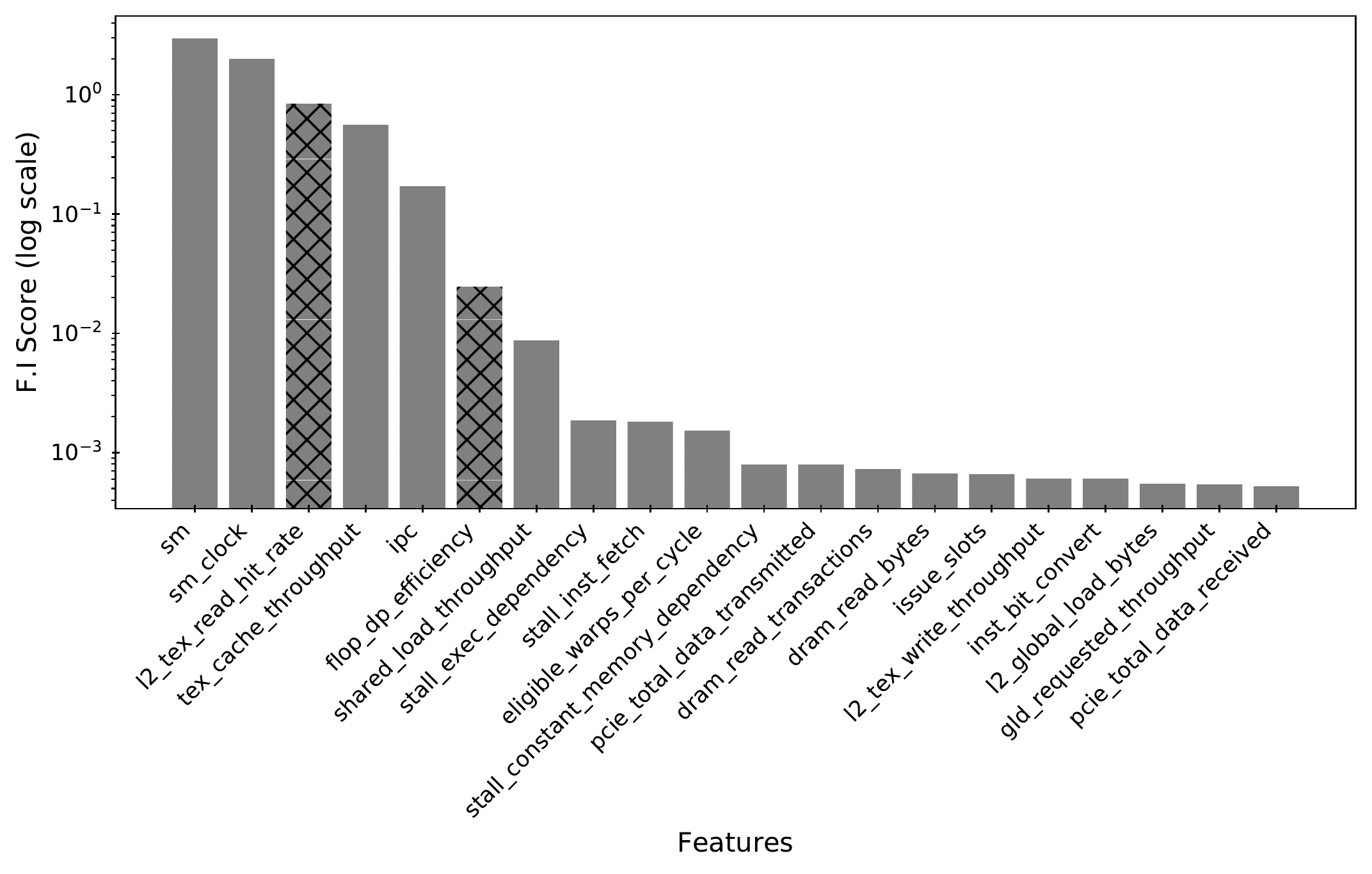}
   \caption{Energy prediction model}
   \label{fig:fa_power}
\end{subfigure}
\begin{subfigure}[t]{0.48\textwidth}
\includegraphics[width=\linewidth]{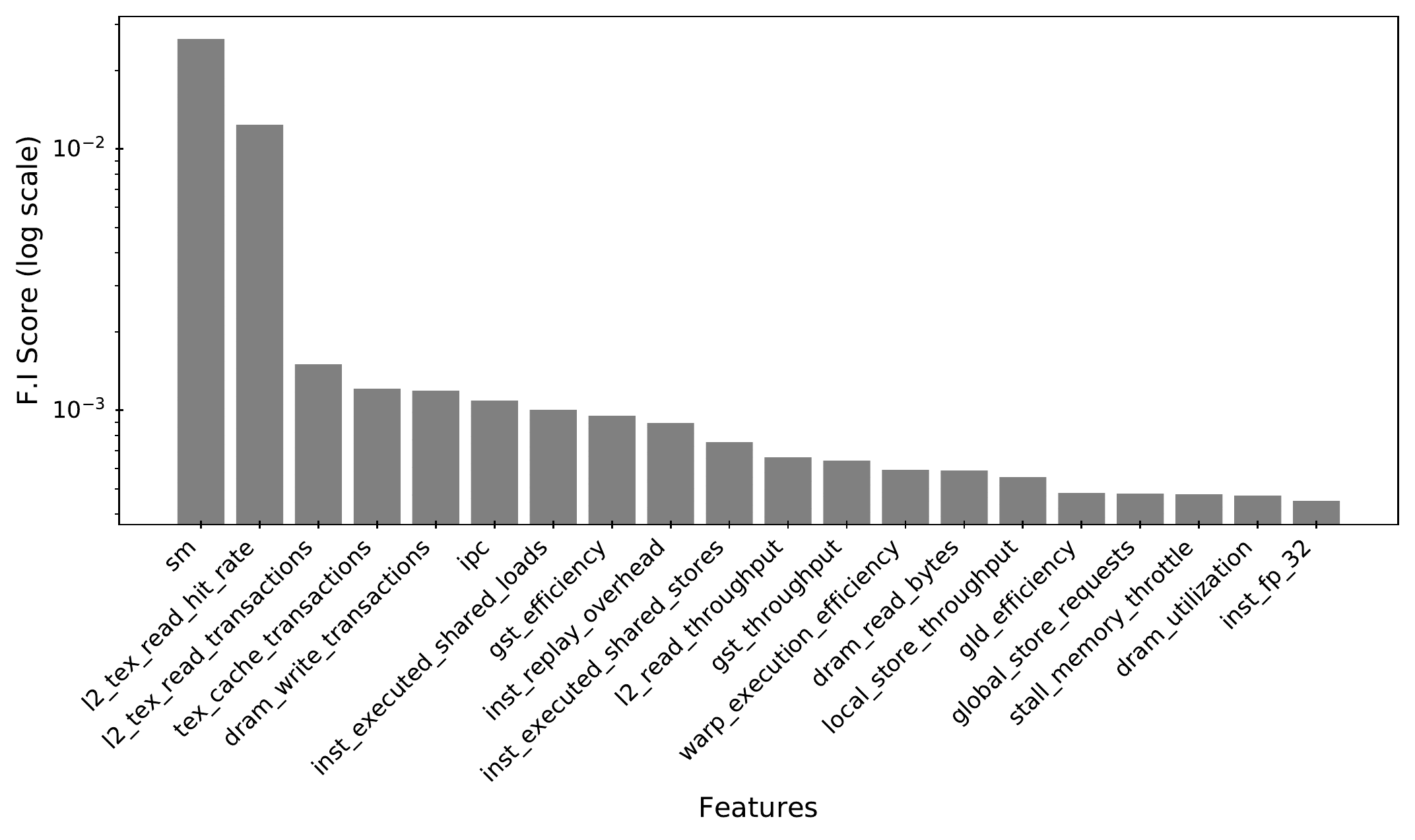} 
   \caption{Time prediction model}
   \label{fig:fa_time}
\end{subfigure}
\caption{Top 20 features sorted based on Feature Importance (F.I) score  (difference in loss value with and without the feature)}
\label{fig:fa-fi-score}
\end{figure*}
\begin{figure}
  \captionsetup{justification=centering}
\includegraphics[width=\linewidth]{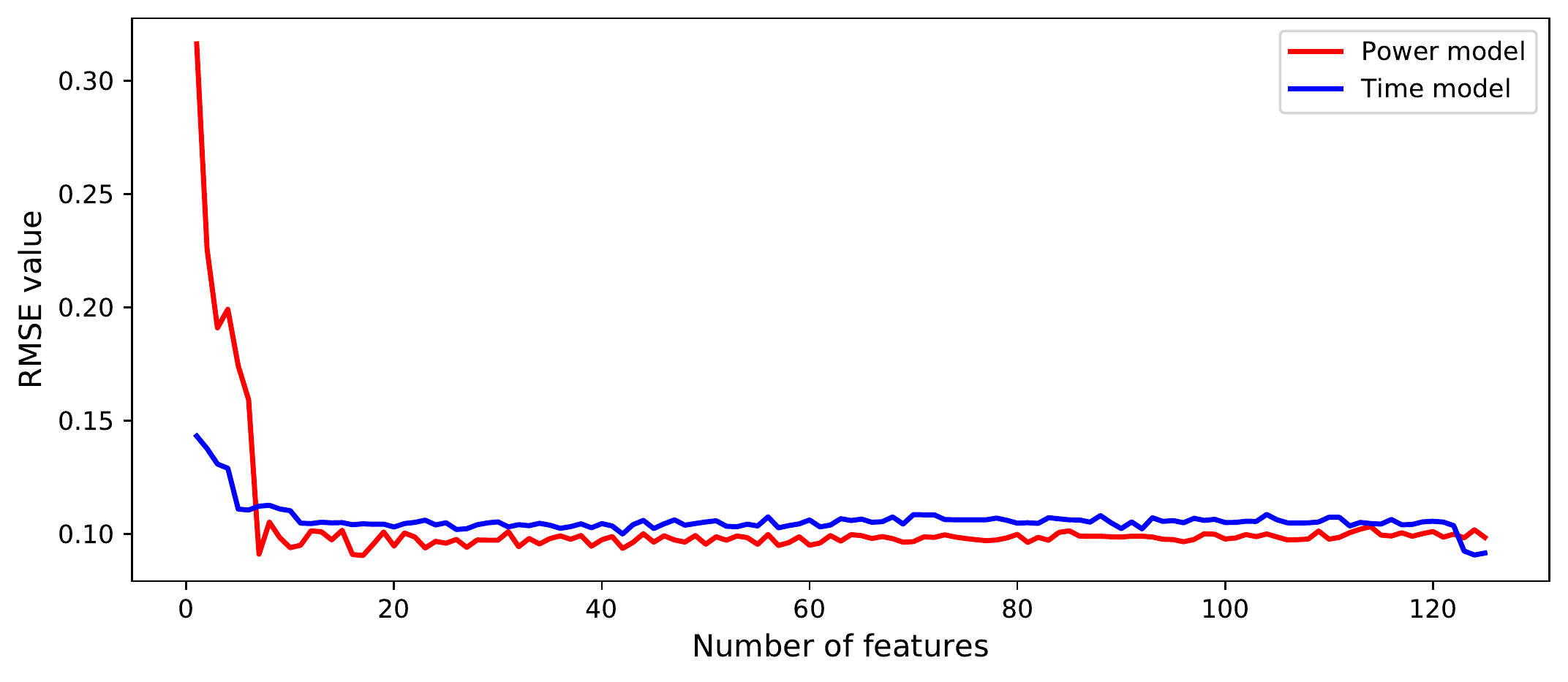} 
   \caption{Threshold analysis of features}
   \label{fig:fa_threshold_analysis}
\end{figure}

We carry feature analysis to understand the importance of individual features towards the performance of the prediction model. It also represents the features that are highly influential on the prediction output. 

Figure \ref{fig:fa-fi-score} indicates the Feature Importance ($F.I$) score of different features. We plot the twenty most significant features sorted in descending order of their score. Here, $F.I$ value represents the difference between the loss value of the model with and without that feature. The model without this feature is similar to the one that would have been trained if this feature was excluded from the data set. Since,  \textit{RMSE} is our loss function, the  F.I score on $y$ axis shows change in \textit{RMSE} value.

We can observe from Figure \ref{fig:fa_power} and Figure  \ref{fig:fa_time}, different types of features contribute to different magnitude while predicting energy and time of application, respectively. In both cases,  feature $sm$, which represents the streaming multiprocessor's utilization, has the highest $F.I$ score showing its high impact on the energy and time. Furthermore,  $sm\_clock$ is the second most important feature in predicting energy, reflecting a direct correlation between energy and frequency clock settings. Please note, since our testbed GPUs (Tesla-P100) have only one memory frequency ($f_{memclock}$), it does not feature in the top twenty features as there is no variation introduced by it in the data set. For the system with multiple $f_{memclock}$, it is expected that it would have a significant effect on the model's performance.  We can also observe from time model in Figure \ref{fig:fa_time}, different features present when compared to the energy model (Figure \ref{fig:fa_power}). A total of 5 features are in common between two models. The features related to l2 cache and stall dependencies have more impact in the energy model, while in time prediction,  separate features like $inst\_executed\_shared\_stores$, $inst\_fp\_32$ have occurred in top 20 features showing the higher co-relation between execution time and the metrics related to instruction count. 

To further analyze the impact of the number of features on the prediction model's performance, we carry a threshold analysis. Here, features are sorted based on their $F.I$ score, and recursively added to the training data set, and resulting loss value (\textit{RMSE}) is calculated accordingly. As shown in the Figure \ref{fig:fa_threshold_analysis}, for both the power and time model, the top 20 features are sufficient enough to achieve the reasonable performance with excellent RMSE  value and further inclusion of features do not yield much improvement in the result without increasing the training cost. 

\begin{table}[]
\centering
\resizebox{\columnwidth}{!}{%
\begin{tabular}{lllll}
\hline
\textbf{Catboost model} & \textbf{$depth$} & \textbf{$l2\_leaf\_reg$} & \textbf{$iterations$} & \textbf{$learning\_rate$}   \\
\hline
Power                  & 4              & 5                  & 1200                & 0.1                      \\
Time                   & 4              & 3                  & 1200                & 0.03    
\\
\hline
\end{tabular}
}
\caption{Optimal parameters obtained for CatBoost from grid search technique}
\label{table:girdsearch}
\end{table}

\subsection{Feature Correlation with Clustering}\label{sec:fcorealtion}
The prediction models need exhaustive application profiling data from multiple frequency combinations. However, profiling every new application is tedious and infeasible in real-time. Thus, using the existing data and correlating with the new application is a common practice in profiling-based predictive modeling scenarios \cite{kriegel2009clustering}. In such a case, a new application should have at least one set of profiling data of one frequency combination; we take the default clock as a reference for this.  We generate the clusters based on already collected exhaustive data and predict the cluster label for a new application. Furthermore, a highly correlated application within the cluster is chosen from simple heuristic, i.e., application with the lowest absolute difference in execution time is selected to further match the applications performance-similarity. 

We use the same set of twelve applications to perform this analysis. To generate the clusters, we use  K-means. We found that an optimal number of clusters is five based on the weighted sum of the squared error metric. Applications belonging to a different group and their correlated application can be seen in Table \ref{table:clustering}. The cluster-3 has only one application, i.e., 2MM, suggesting the essence of having a still more number of applications in the sample space to have at least two or more applications in each cluster. 
The robustness of this method is evaluated by predicting the execution time and energy for all the applications using the profile data of corresponding correlated applications. The average RMSE value of 3.19 and 1.11 is achieved for energy and time prediction, respectively, proving the feasibility of this method.   
\begin{table}[]
\begin{tabular}{lll}
\hline
\textbf{Applications} & \textbf{Cluster label} & \textbf{Correlated  application} \\
\hline
particlefilter\_naive & 0                      & particlefilter\_float            \\
particlefilter\_float & 0                      & particlefilter\_naive            \\
myocyte               & 1                      & lavaMD                           \\
lavaMD                & 1                      & myocyte                          \\
Backprop              & 2                      & ATAX                             \\
SYRK                  & 0                      & particlefilter\_float            \\
SYR2K                 & 0                      & particlefilter\_naive            \\
GEMM                  & 4                      & CORR                             \\
COVAR                 & 4                      & CORR                             \\
CORR                  & 4                      & COVAR                            \\
ATAX                  & 2                      & Backprop                         \\
2MM                   & 3                      & 2MM       \\
\hline
\end{tabular}
\caption{Cluster labels and correlated app}
\label{table:clustering}
\end{table}

\section{Deadline Aware Application Scheduling by Data-Driven DVFS}\label{Section:data-driven-dvfs}
The advantages of power and performance estimations of  GPU workloads are manifold. It is used in resource management techniques like scheduling \cite{capodieci2018deadline}, power capping\cite{komoda2013power}, and also in the analysis of performance bottlenecks of workloads \cite{ge2013effects}. In this work, we propose deadline aware energy-efficient application scheduling guided by the data-driven DVFS.

\begin{figure}
  \captionsetup{justification=centering}
\begin{subfigure}[t]{0.24\textwidth}
\includegraphics[width=\linewidth]{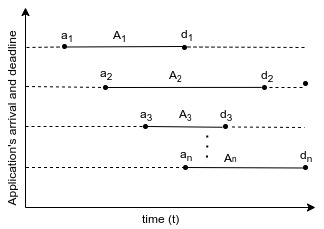}
   \caption{Workload Model}
   \label{fig:workloadmodel}
\end{subfigure}
\begin{subfigure}[t]{0.24\textwidth}
\includegraphics[width=\linewidth]{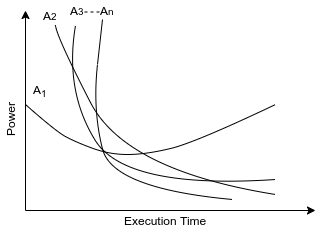}
   \caption{Power and Execution time}
   \label{fig:powercurve}
\end{subfigure}
\caption{Workload and Power-Execution time models}
   \label{fig:powercurve}
\end{figure}

 The workload model of our scheduling is shown in Figure \ref{fig:workloadmodel}. It  consists of a set of applications represented as a vector $W = \{A_1, A_2, A_3.. A_n \}$, with their own arrival and deadline times, represented as vectors  $ A= \{a_1, a_2, a_3, ..., a_n\}$ and  $ D= \{d_1, d_2, d_3, ..., d_n\}$ where $ \forall a_i \in A $ and  $ \forall d_i \in D $, respectively. 

As illustrated in Figure \ref{fig:powercurve}, the power curve for individual applications are non-linear with their execution time. The objective is to configure the frequency that meets application $A_i$'s deadline $d_{i}$ and also has the lowest energy consumption according to it's power curve. Therefore, considering the energy consumption of application $i$ is $P_{i}$, then 
reducing the overall GPU energy consumption is formulated as a minimisation problem as below:
\begin{equation}
\label{eq:problemformulation}
\begin{aligned}
    & \underset{}{\text{minimize}}
    & & P_{total} = \sum_i^n P_{i}  \\
    & \text{subject to}
    && \forall T_{i}  \leq  d_{i} \\
\end{aligned}
\end{equation}
In the above Equation \ref{eq:problemformulation}, $T_{i}$ is application's execution time. The objective function minimises the total GPU energy consumption and the constraint makes sure that application is executed within the  deadline. However, solving Equation \ref{eq:problemformulation} is equivalent to constrained global optimization which is  an NP-hard problem \cite{pardalos1987constrained}. As it is impractical to find an optimal solution in real-time, we present a heuristic algorithm in order to reduce the problem complexity suitable for on-line scheduling and find a near-optimal solution within reasonable amount of time. 

\begin{algorithm}[t]
\caption{Deadline-aware  Scheduling by Data-Driven DVFS}
\label{alg:scheduling}
\begin{algorithmic} [1]
\Statex \textbf{Inputs:}
\STATE $W:$ list of  applications to be executed (workload)
\STATE $D:$ list of  application's deadline (d)
\Statex \textbf{Output:} GPU application clock and schedule
\WHILE {true}
    \STATE jobQue  $\leftarrow$ \textsc{getCurrentArrivalJobs($W$)}
    \STATE jobQue  $\leftarrow$ \textsc{sort}(jobQue, key$=$ $D$, order$=$ asc)
    \STATE clockList  $\leftarrow$ \textsc{getGPUSupportedCloks}(deviceID)
    \FOR{\textbf{each} $job$  in jobQue}
        \STATE predictionInput $\leftarrow$ \textsc{GetCorrelatedData}($job$) 
        \STATE minPower $\leftarrow$ MAX
        \STATE maxTime $\leftarrow$  \textit{job.$d$}
        
        \STATE  ($f_{smclock}$, $f_{memclock}$) $\leftarrow$ NULL
        \FOR{\textbf{each} $clockSet$  in clockList}

               \STATE $\hat{P_{job}}$ $\leftarrow$  \textsc{predictPwr}(predictionInput.$clockSet$)
               \STATE  $\hat{T_{job}}$ $\leftarrow$  \textsc{predictTime}(predictionInput.$clockSet$)
                \IF{ $\hat{P_{job}}$ $<$ minPower \textbf{and}  $\hat{T_{job}}$ $<$ maxTime }
                    \STATE minPower $\leftarrow$   $\hat{P_{job}}$
                    \STATE maxTime $\leftarrow$   $\hat{T_{job}}$
                    \STATE ($f_{smclock}$, $f_{memclock}$) $\leftarrow$  $clockSet$ 
                \ENDIF
        \ENDFOR
        \STATE  \textsc{setGPUApplicationclock}($f_{smclock}$, $f_{memclock}$)
        
        \IF{ \textbf{not} ($f_{smclock}$, $f_{memclock}$)  NULL}
        \STATE  \textsc{execute}(job)
        \ENDIF
    \ENDFOR{}
\ENDWHILE
\Statex \textbf{End}
\end{algorithmic}
\end{algorithm}

The Algorithm \ref{alg:scheduling} shows proposed scheduling algorithm for deadline-aware application scheduling with the data-driven DVFS.
Its objective is to  reduce the energy  while meeting the application's deadline and it is achieved by generating an efficient schedule sequence and also suitably scaling the GPU frequencies. The algorithm takes the application list, and their corresponding deadlines as an input and outputs suitable predicted clock and scheduling sequence for applications.

First,  according to applications arrival time, the available jobs are sorted based on the deadlines in ascending order (line 4,5) to make sure the jobs with the earliest deadline are executed first. Considering the new arrived application has only default clock input profile data,  we find its correlated application and use it's exhaustive profile data for prediction (as explained in Section \ref{sec:fcorealtion}). Furthermore, the power and execution time is predicted for all the supported GPU frequency clock sets (lines 12-14) using the prediction models proposed in Section \ref{Section:data-driven-dvfs}. For a given job, the clocks which have the lowest power consumption and also the predicted execution time less than its deadline is selected (15-18). Finally, the selected application clock is configured, and the application is executed. 

The time complexity of Algorithm \ref{alg:scheduling} is polynomial. Assuming we have $n$ jobs to be scheduled on a GPU with $c$ number of clocks. The  sorting of jobs  require worst-case time complexity of $n\log n$ (line 5). Furthermore, each job has to be evaluated on all clock-sets and executed, that has a time complexity of  $nc$ (line 7-20). Hence, the total complexity will be $(n\log n + nc)$. Considering $c$ is a constant for any given GPU, the Algorithm \ref{alg:scheduling}  has polynomial  time complexity of  $O(nlogn)$.

\section{Performance Evaluation}
We discuss the implementation of our proposed algorithm integrated with the prediction models. We also analyze and discuss the results compared to baseline algorithms.
\subsection{Implementation}
We implemented the proposed scheduling framework using Python language. We developed a multithreaded application where the main thread executes the algorithm \ref{alg:scheduling}'s logic and invokes the application execution files. Additionally, it also launches two other background threads, one to collect the GPU data related to energy metrics (by running bash scripts with $nvidia-smi$ $dmon$) and other kills the background thread once the application execution is done. Furthermore, the application clocks are predicted based on the proposed model in Section \ref{Section:data-driven-dvfs}, and these predicted clocks are set before executing the scheduled application using the NVML's $nvidia-smi$'s API. 

\subsection{Experimental Setup}
We use  Grid'5000  testbed \cite{Grid5000} for our experiments. It is a large-scale flexible testbed for experiment-driven research, specifically,  designed for experimental evaluations of the energy-efficient techniques \cite{Grid5000}. We have used  Tesla-P100 GPUs for our experiments. 
This machine has a dual CPU Intel Xeon Gold 6126 processor with 12 cores per CPU and 192 GB of primary memory. The GPU itself has 
3584 cores with 16 GB primary memory. 
The machines are deployed with Debian 10 as the operating system installed with CUDA 10.1 drivers. 

\subsection{Benchmarking Applications} We have used twelve applications from two bench-marking suits (PolyBench and Rodinia), which are also part of our predictive modeling (Table \ref{table:benchmark}). To formally produce the workload model described in Figure \ref{fig:workloadmodel},  initially, we use  GPU default application clocks run time ([715, 1189] MHz for Tesla-P100)   as a reference to our application's execution time.  
Also, the application's arrival time and deadline are calculated based on the normal distribution. For the arrival time, the minimum and maximum value range of distribution are set to (1, 50), and for the deadline, it is set to (1, 2$\times$ execution time). This is to make sure that the application's deadline can have a maximum value of twice their execution time. All these applications are CUDA-based implementation, and configurations are  shown in Table \ref{table:benchmark}.

\subsection{Analysis of Results}
We evaluate the proposed algorithm \ref{alg:scheduling} against two baselines. \textit{(1):} \textit{Default Clock (DC)}: GPU frequencies  are set to default application clocks. The applications usually run on default clocks under normal conditions \textit{(2):} \textit{Max Clock (MC)}: GPU frequencies are set to maximum possible frequency domains. This is a widely used technique in the form of near-threshold computation or computational sprinting to finish the execution as fast as possible under strict performance requirements \cite{Basu2016}. We represent our proposed policy as D-DVFS, data-driven DVFS.
  \begin{figure}
\includegraphics[width=\linewidth]{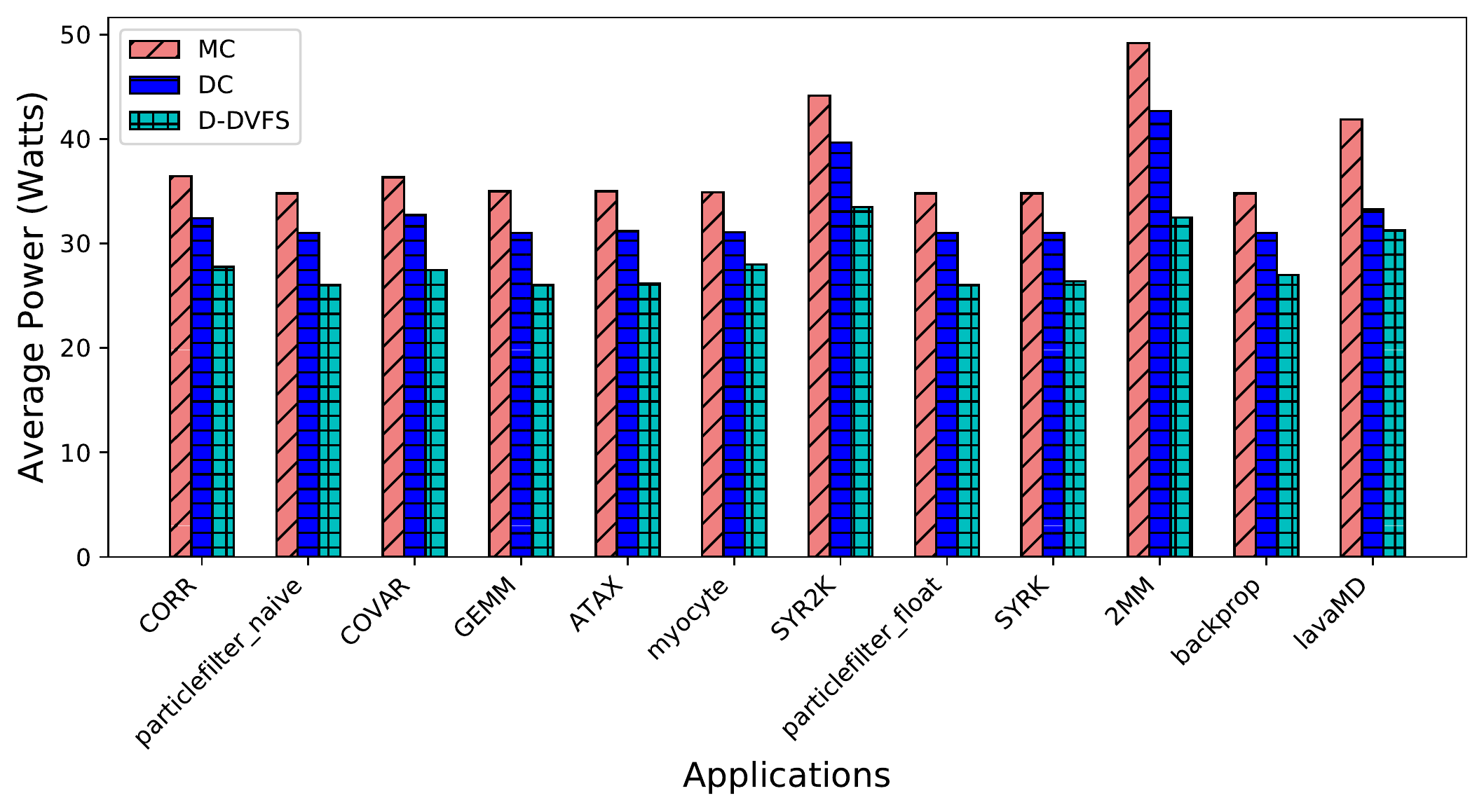}
   \caption{Average  energy consumption of applications}
   \label{fig:application_power}
\end{figure}

 \begin{figure}
\includegraphics[width=\linewidth]{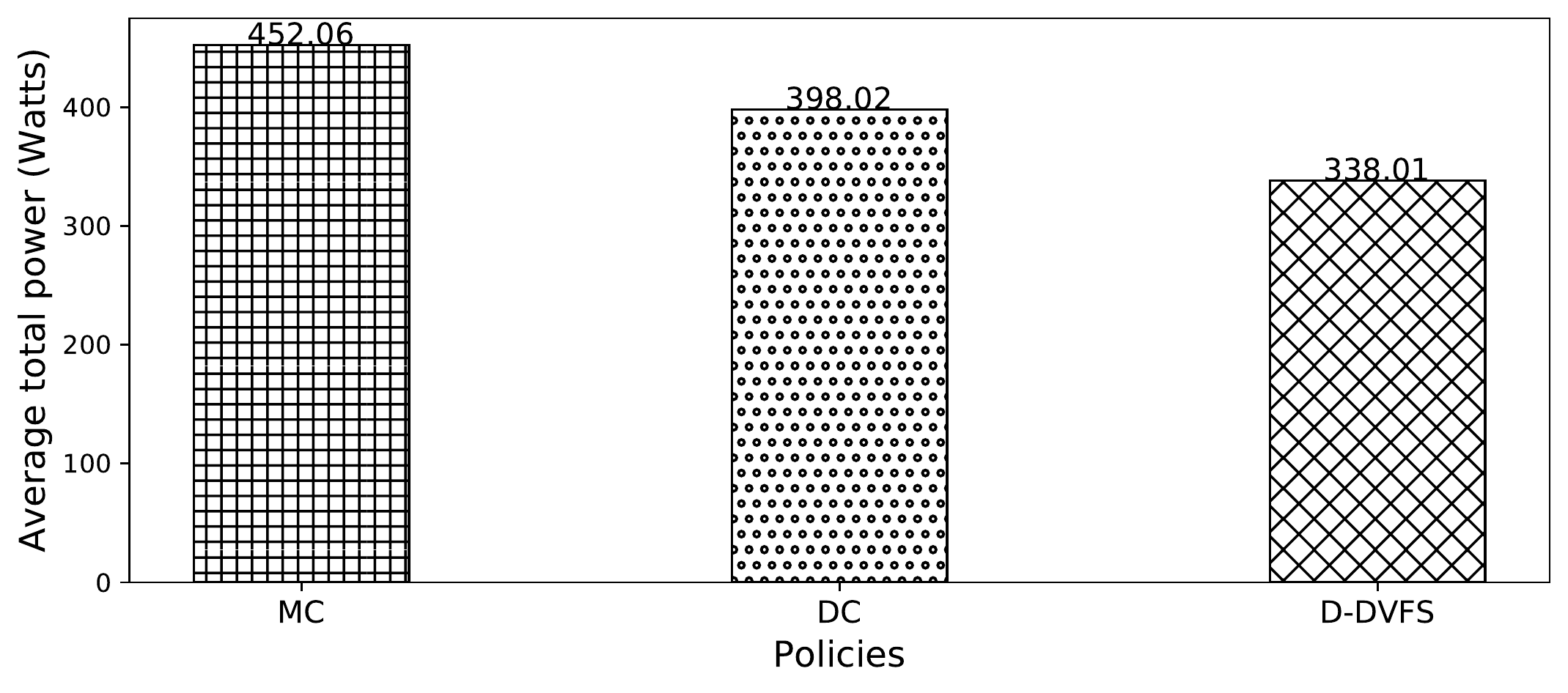}
 \caption{Average total energy consumption  of GPU}   \label{fig:total_power}
\end{figure}

Figure \ref{fig:application_power} depicts  energy consumption of various applications by different policies. Both the MC and DC policies consume a much higher amount of energy as compared to our proposed data-driven frequency scaling (D-DVFS). Particularly, the MC  consumes more energy than the other two policies. Since D-DVFS sensibly configures the clocks to the lowest possible energy consumption, it leads to significant energy savings for the application.

The total average GPU energy consumption can be seen in Figure \ref{fig:total_power}. MC, DC, and D-DVFS have an average of 452.06 (W.S), 392.02 (W.S), and 338.01 (W.S), respectively. In other words, D-DVFS consumes 15.07\% and 25.3\% less than MC and DC policies, respectively. The results confirms that  D-DVFS selects energy-efficient frequencies for application execution.

\begin{figure}
\includegraphics[width=\linewidth]{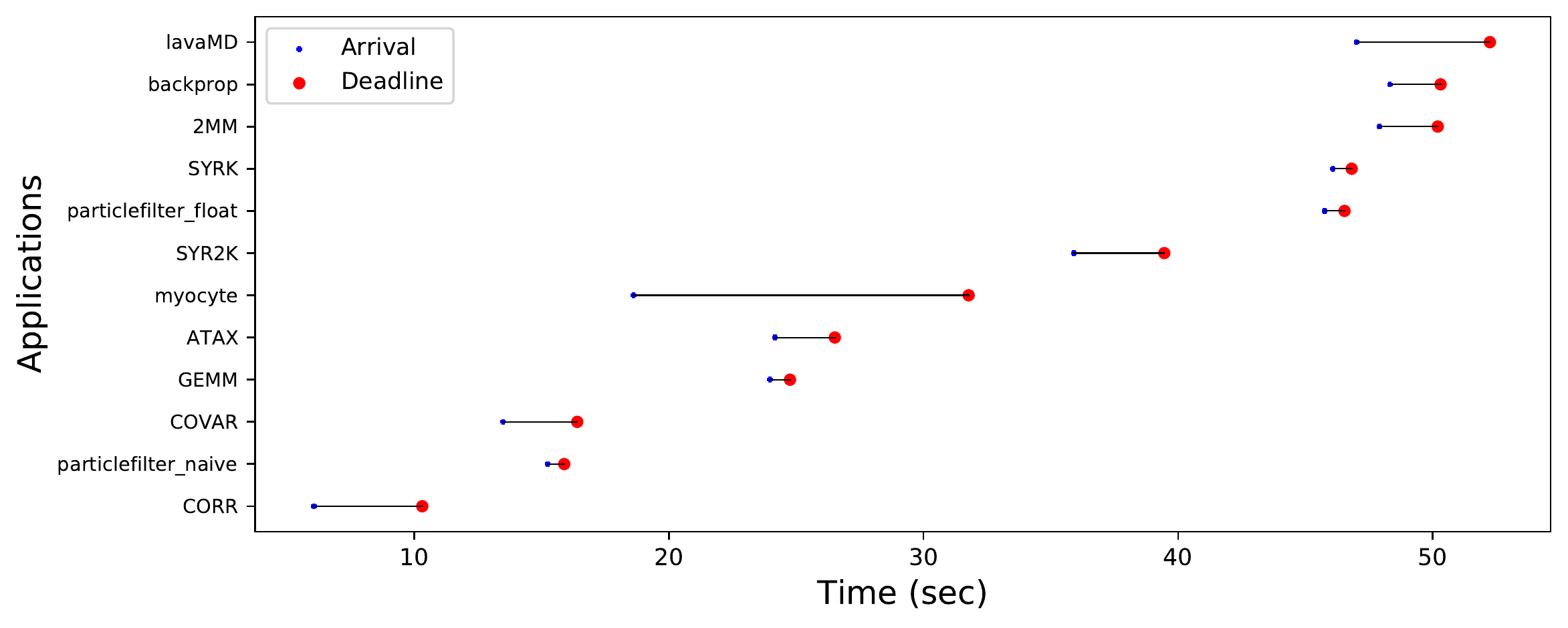}
  \caption{Application arrival and deadline times}
  \label{fig:arrivaldeadline}
\end{figure}

\begin{figure}
\includegraphics[width=\linewidth]{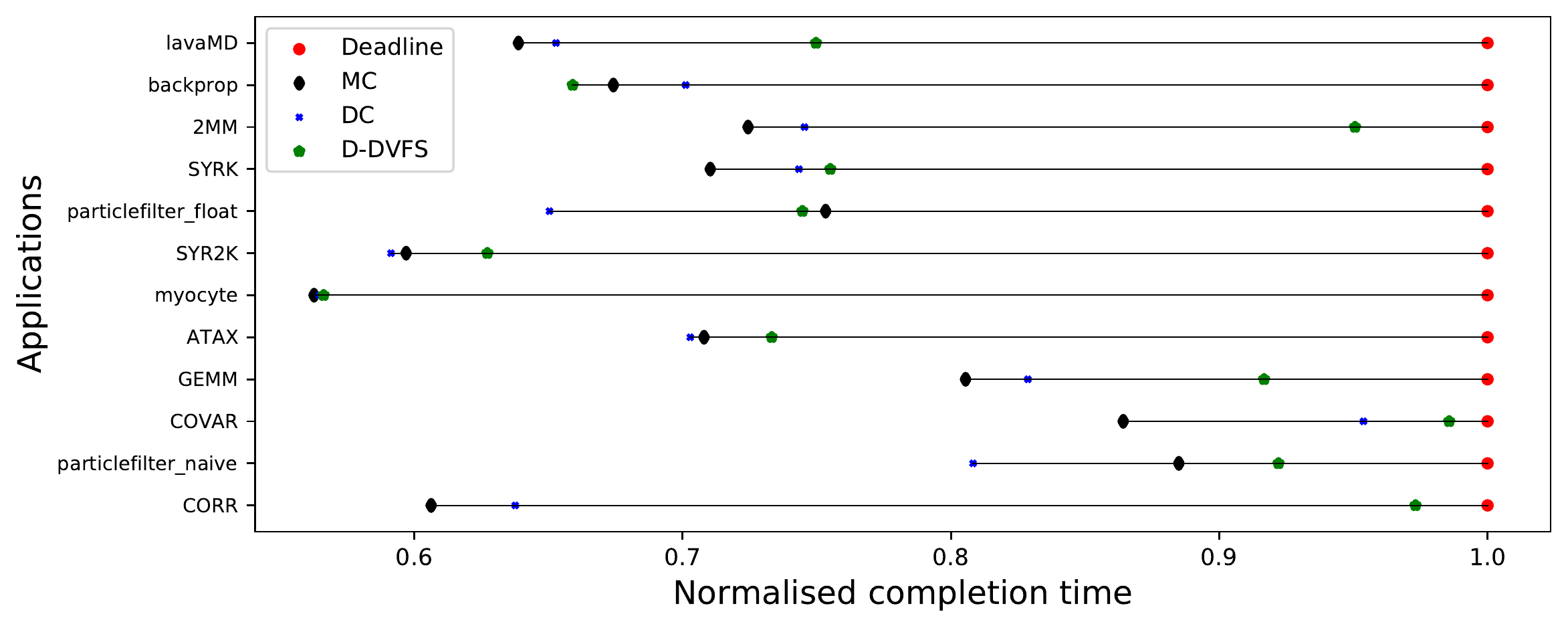}
  \caption{Normalised application completion time compared to deadline}
  \label{fig:deadline}
\end{figure}

The application's arrival and deadline time generated using distribution are shown in Figure \ref{fig:arrivaldeadline}. Accordingly,  the normalized application completion time achieved using our scheduling and baseline policies is shown in  Figure \ref{fig:deadline}. The  D-DVFS policy meets all the deadlines. It tends to execute near to the deadline requirement of applications as it scales to the frequency that has high energy efficiency and predicted execution time that meets the deadline. Although DC and MC policies execute faster, their deadline-agnostic nature tends to run the applications with high frequency and thus spending more energy.
Furthermore, D-DVFS with much lower frequency executes faster in a few scenarios (for application backprop and particle\_float, refer Figure \ref{fig:deadline}), which represents, faster execution of applications do not have high frequency all the time. Such condition usually happens when the application has significant  I/O  wait or dependency stalls and setting the higher frequency in such scenarios wastes more power. Nevertheless,  our approach inherently learns such behavior and adapts accordingly.

\begin{figure}
\includegraphics[width=\linewidth]{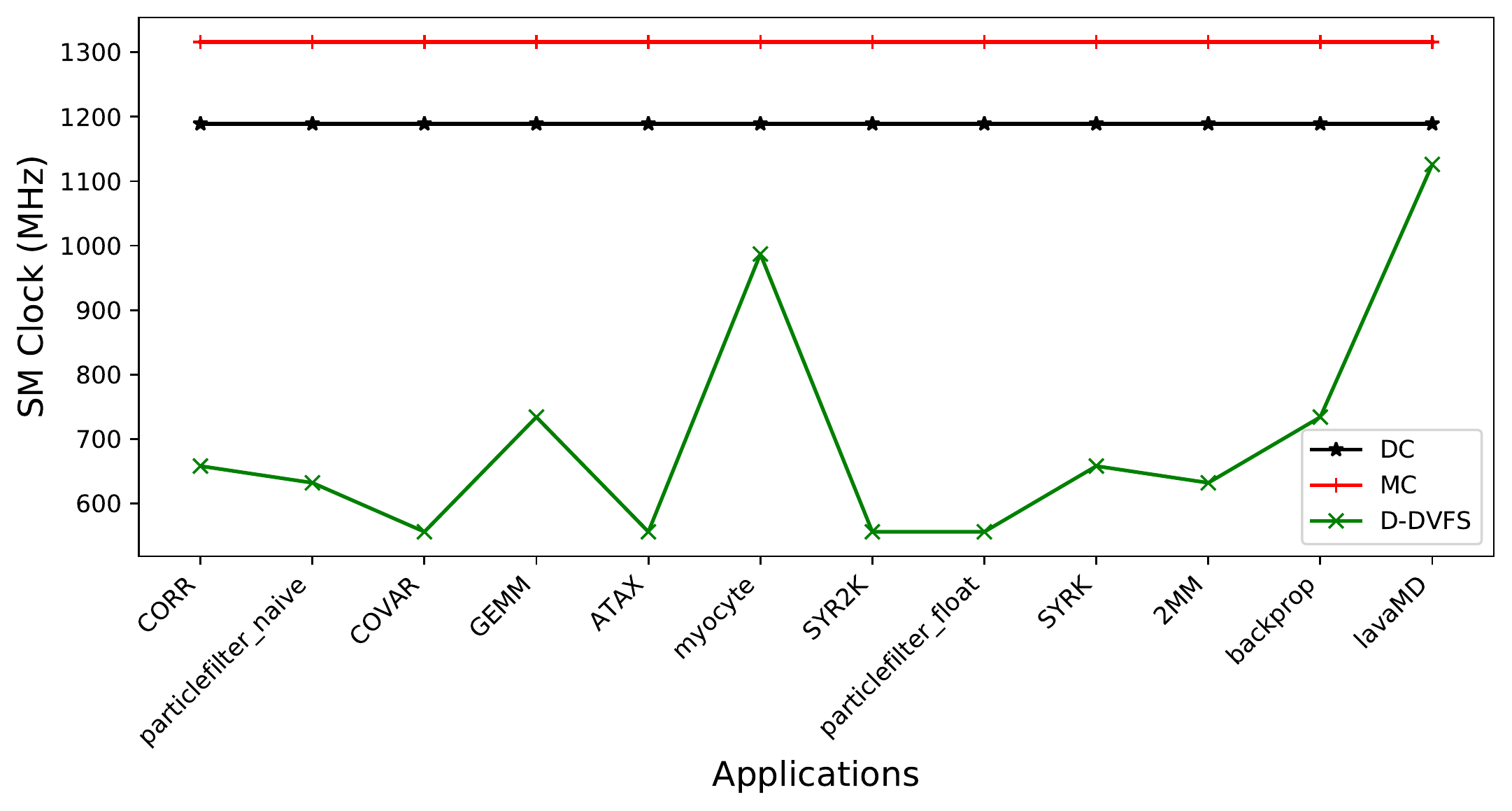} 
  \caption{Frequency Scaling  by different  policies}
  \label{fig:dvfs_scaling}
\end{figure}

The GPU   frequency scaling behavior for different applications is depicted in Figure \ref{fig:dvfs_scaling}. Here, MC and DC always operate statically with the highest possible and default frequencies of the GPU. However, D-DVFS selects much lower frequencies for most of the applications, which are sufficient enough to meet the deadlines. Moreover, for applications that demand faster execution to meet their deadlines, it appropriately scales the frequency and chooses the efficient higher frequency range, this can be evidenced in Figure \ref{fig:dvfs_scaling} for the applications \textit{lavaMD} and \textit{myocyte}.

\begin{figure}
  \captionsetup{justification=centering}
  \begin{subfigure}[t]{0.48\linewidth}
\includegraphics[width=\linewidth]{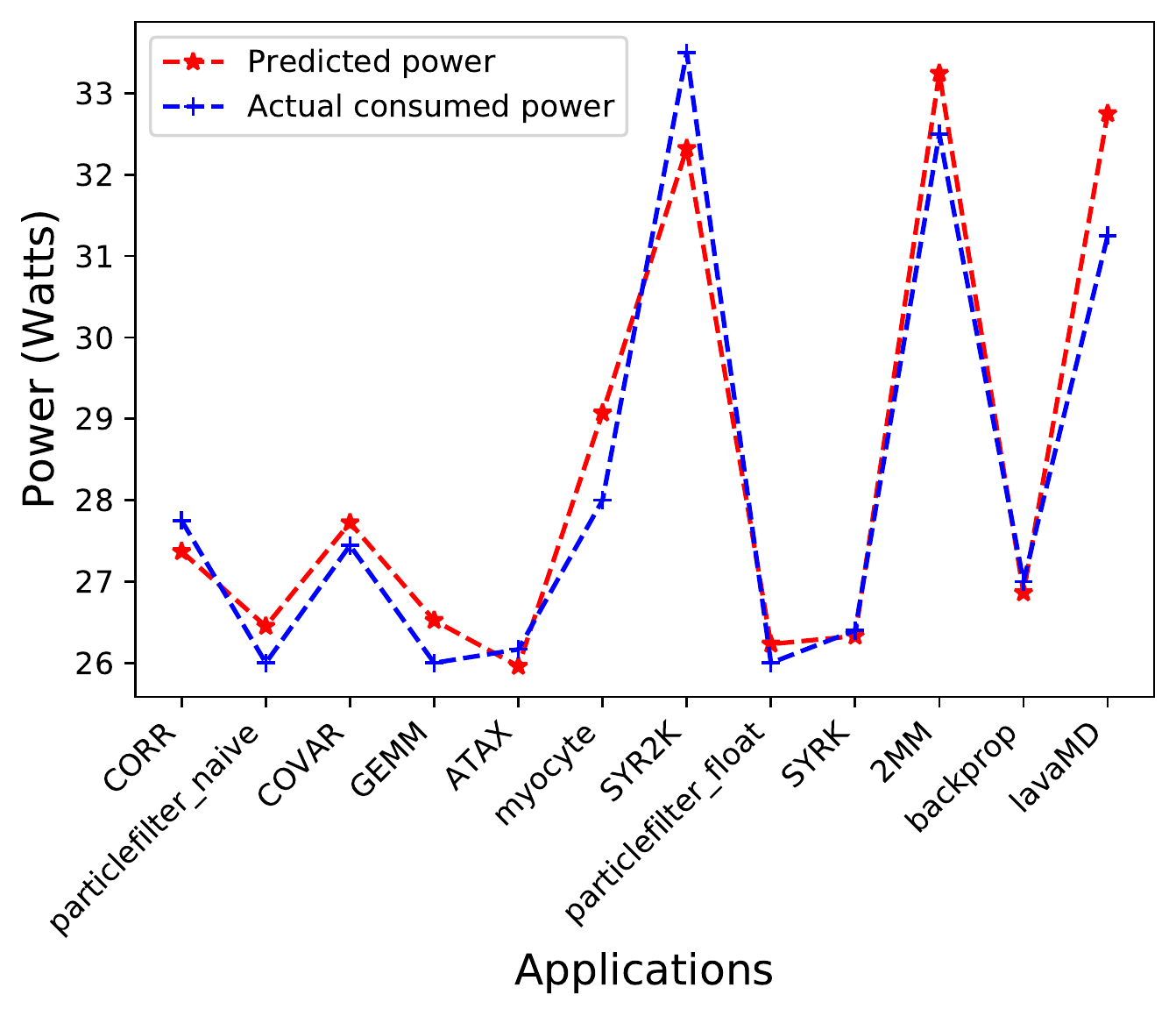}
   \caption{Power}
   \label{fig:predict_power}
\end{subfigure}
\begin{subfigure}[t]{0.48\linewidth}
\includegraphics[width=\linewidth]{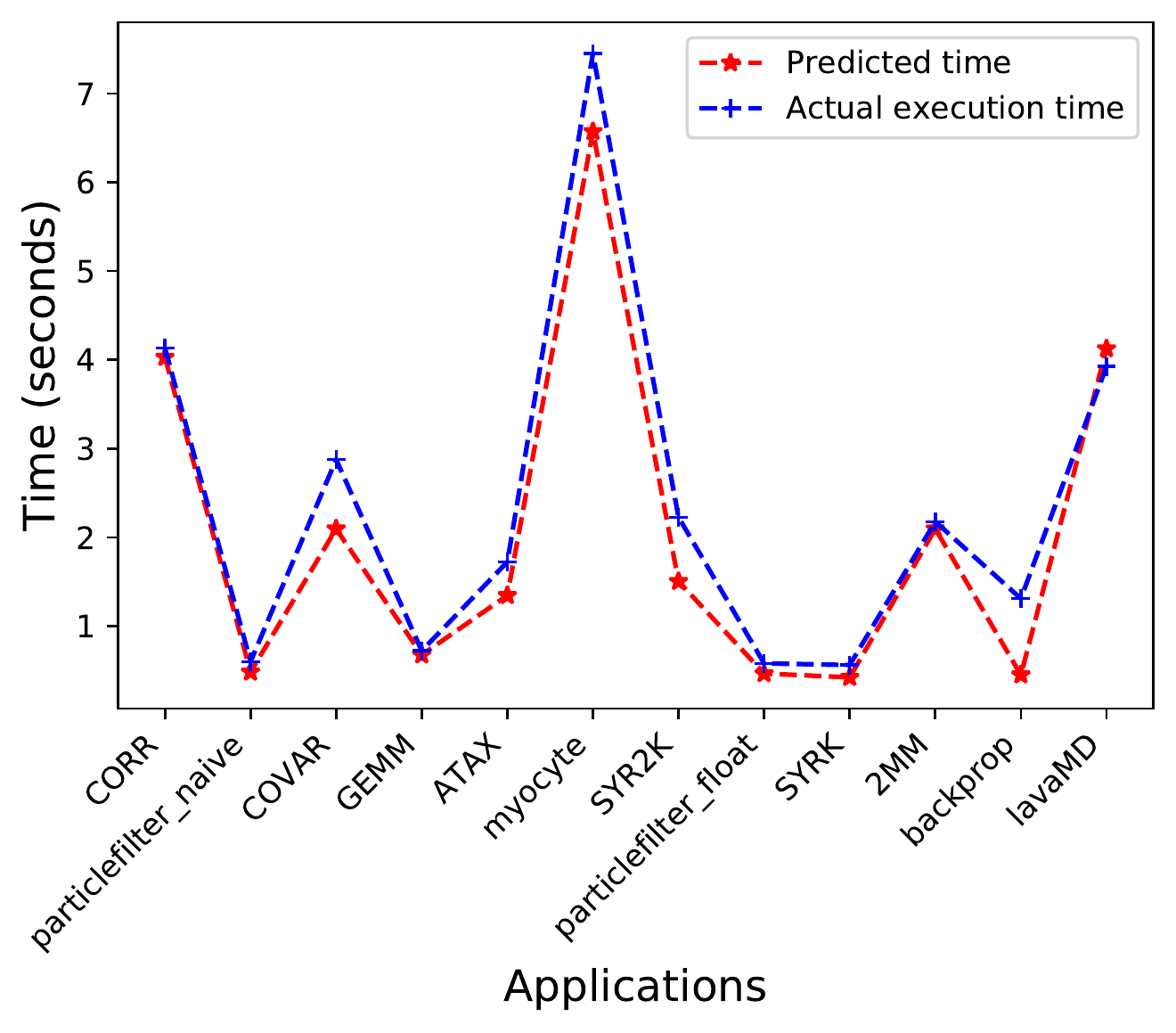}
   \caption{Time}
   \label{fig:predict_time}
\end{subfigure}
\caption{Actual and prediction values in scheduling}
\label{fig:predict_actual}
\end{figure}

The accuracy of prediction models in the scheduler is vital for achieving the stated objective. The performance of energy and time prediction models is shown in Figure \ref{fig:predict_actual}. The predicted values closely follow the actual measurements from the executions showing the accuracy of predictions and thus assisting the scheduler efficiently for frequency scaling.

Therefore, the optimal configuration of frequencies is vital to reduce GPU energy consumption. It is more necessary when different applications have different deadlines. This is the most real case where multiple users submit parallel GPU jobs with their expected deadlines (in the form of wall-time in HPC environments). Employing such techniques, provided they have single execution profiled data, will benefit primarily to save the system energy and also provide better service for application users. 

\section{Related Work}
Several researchers have studied a different aspect of GPU DVFS optimization. 
The existing GPU frequency and performance estimation models can be classified into three types. First, the analytical models \cite{Teo2015} \cite{Chau2017}, \cite{Ma2012}, which uses the mathematical relationships between different system components and workload characteristics. Second,  static models \cite{Wang2019, Fan2019}, usually constructed using source code level metrics or static hardware specifications. Finally,  machine learning models \cite{wu2015gpgpu, guerreiro2019dvfs}, where different predictive models are employed to estimate the required parameters accurately. 

Losch et al. \cite{Losch2018} present an accurate analytical energy model for a task execution on heterogeneous nodes by characterizing the application execution and energy model.
Some works have also explored techniques like power capping and scheduling \cite{komoda2013power, Chau2017} for energy optimization using DVFS and task mapping. The authors in \cite{komoda2013power} have used empirical, analytical models to configure the CPU-GPU frequency to execute applications within a  power budget. Chau et al. \cite{Chau2017} have studied energy-efficient job scheduling in CPU-GPU systems by regulating the DVFS. The authors proposed analytical approximation algorithms with linear programming (ILP) model and introduced a  heuristic algorithm to solve this problem.  Although analytical models are fast, they fail to accurately estimate the intended metrics due to their sensitivity to different parameters involved in the modeling. 

The static models rely mostly on the source-code or compiler level metrics to build the models. Wang et al. \cite{Wang2019}  proposed a hybrid framework for fast and accurate GPU performance estimation through source-level analysis. They used a total of 23 parameters collected from the hardware specifications, simulation traces, and the source code.
Fan et al. \cite{Fan2019} also studied predicting the energy and performance using the static source code features from several real and synthetic open-CL kernels. Although their prediction model relies on ML techniques (Support vector regression- SVR), the training data is collected from the static source code features.  They use Pareto-set of frequency configurations to find the optimal scaling values between speedup and energy further. However, models built using static features perform poorly when applied to different  GPU architectures as each device has a different response to the energy and execution time. Therefore, it is beneficial to build models with actual data from the real-platforms.

Machine Learning (ML) models have been used by researchers recently in GPUs DVFS management.  Wu et al. \cite{wu2015gpgpu} proposed a neural network model to estimate the scaling curve of application with different hardware configurations. While their objective is tuning different hardware parameters, we instead focus more on configuring the frequency domains to facilitate the efficient DVFS for application execution. Similarly, Guerreiro et al. \cite{guerreiro2019dvfs}, investigated the DVFS-aware application classification to improve GPU efficiency. They characterize the applications using the nine different application profiling features and classify the workloads based on the hierarchical clustering and neural network classifier. Our approach is different where we predict the energy and time with varying settings of the clock while this method classifies application into different domains and optimize accordingly.
Furthermore, Tang et al. \cite{Tang2019} carried out an empirical study of GPU DVFS on energy and performance of deep learning workloads. They analyze the effect of DVFS with different core frequencies while training the deep neural networks on  NVIDIA  $GTX 2080T_i$. The empirical results have shown that optimal frequency settings can significantly save energy consumption.
Most of these works focus on a single objective. However, in this work, we propose a data-driven frequency scaling approach for the deadline-aware scheduling algorithm.

\section{Conclusions and Future Work}
Optimal configurations of GPU frequencies can significantly reduce energy consumption. However, identifying the suitable frequencies that result in lower energy consumption with the strict application's deadline requirement is extremely challenging. This is mainly due to the complexity induced by the application's response to energy, execution time, and clock settings. To that end, we present a framework that selects suitable GPU frequencies for a given application using the data-driven techniques and accordingly schedule the applications while reducing energy consumption and meeting  deadline. Our model achieves high accuracy with average RMSE values of 0.38 and 0.05 for energy and time, respectively, indicating that predicting the energy is quite difficult as compared to the execution time. Additionally, our proposed scheduling algorithm consumes 15.07\% less energy as compared to the baselines while satisfying the deadline requirements. 

In the future, we plan to extend this framework to multi-GPU platforms. We also intend to consider deep learning workloads where frequency tuning can achieve more benefits due to their long-running nature. 

\section*{Acknowledgment}
We thank Laurent Lefevre and  Marcos Assuncao at INRIA, France, for granting access to  Grid'5000 testbeds. 
\bibliographystyle{IEEEtran}

\bibliography{main}

\begin{thebibliography}{10}
\providecommand{\url}[1]{#1}
\csname url@samestyle\endcsname
\providecommand{\newblock}{\relax}
\providecommand{\bibinfo}[2]{#2}
\providecommand{\BIBentrySTDinterwordspacing}{\spaceskip=0pt\relax}
\providecommand{\BIBentryALTinterwordstretchfactor}{4}
\providecommand{\BIBentryALTinterwordspacing}{\spaceskip=\fontdimen2\font plus
\BIBentryALTinterwordstretchfactor\fontdimen3\font minus
  \fontdimen4\font\relax}
\providecommand{\BIBforeignlanguage}[2]{{%
\expandafter\ifx\csname l@#1\endcsname\relax
\typeout{** WARNING: IEEEtran.bst: No hyphenation pattern has been}%
\typeout{** loaded for the language `#1'. Using the pattern for}%
\typeout{** the default language instead.}%
\else
\language=\csname l@#1\endcsname
\fi
#2}}
\providecommand{\BIBdecl}{\relax}
\BIBdecl

\bibitem{luebke2006gpgpu}
D.~Luebke, M.~Harris, N.~Govindaraju, A.~Lefohn, M.~Houston, J.~Owens,
  M.~Segal, M.~Papakipos, and I.~Buck, ``Gpgpu: general-purpose computation on
  graphics hardware,'' in \emph{Proceedings of the 2006 ACM/IEEE conference on
  Supercomputing}.\hskip 1em plus 0.5em minus 0.4em\relax ACM, 2006, p. 208.

\bibitem{jeon2018multi}
M.~Jeon, S.~Venkataraman, J.~Qian, A.~Phanishayee, W.~Xiao, and F.~Yang,
  ``Multi-tenant gpu clusters for deep learning workloads: Analysis and
  implications,'' Technical report, Microsoft Research, 2018., Tech. Rep.,
  2018.

\bibitem{top500}
\BIBentryALTinterwordspacing
top500, ``top500 supercomputer list,'' 2019. [Online]. Available:
  \url{https://www.top500.org/lists/2019/11/}
\BIBentrySTDinterwordspacing

\bibitem{keckler2011gpus}
S.~W. Keckler, W.~J. Dally, B.~Khailany, M.~Garland, and D.~Glasco, ``Gpus and
  the future of parallel computing,'' \emph{IEEE Micro}, 2011.

\bibitem{exposito2013general}
R.~R. Exp{\'o}sito, G.~L. Taboada, S.~Ramos, J.~Touri{\~n}o, and R.~Doallo,
  ``General-purpose computation on gpus for high performance cloud computing,''
  \emph{Concurrency and Computation: Practice and Experience}, vol.~25, no.~12,
  pp. 1628--1642, 2013.

\bibitem{ilager2019gpu}
S.~Ilager, R.~Wankar, R.~Kune, and R.~Buyya, ``Gpu paas computation model in
  aneka cloud computing environments,'' \emph{Smart Data: State-of-the-Art
  Perspectives in Computing and Applications}, p.~19, 2019.

\bibitem{bridges2016understanding}
R.~A. Bridges, N.~Imam, and T.~M. Mintz, ``Understanding gpu power: A survey of
  profiling, modeling, and simulation methods,'' \emph{ACM Computing Surveys
  (CSUR)}, vol.~49, no.~3, p.~41, 2016.

\bibitem{ebrahimi2014review}
K.~Ebrahimi, G.~F. Jones, and A.~S. Fleischer, ``A review of data center
  cooling technology, operating conditions and the corresponding low-grade
  waste heat recovery opportunities,'' \emph{Renewable and Sustainable Energy
  Reviews}, vol.~31, pp. 622--638, 2014.

\bibitem{mei2013measurement}
X.~Mei, L.~S. Yung, K.~Zhao, and X.~Chu, ``A measurement study of gpu dvfs on
  energy conservation,'' in \emph{Proceedings of the Workshop on Power-Aware
  Computing and Systems}.\hskip 1em plus 0.5em minus 0.4em\relax ACM, 2013.

\bibitem{ge2013effects}
R.~Ge, R.~Vogt, J.~Majumder, A.~Alam, M.~Burtscher, and Z.~Zong, ``Effects of
  dynamic voltage and frequency scaling on a k20 gpu,'' in \emph{2013 42nd
  International Conference on Parallel Processing}, 2013.

\bibitem{guerreiro2019modeling}
J.~Guerreiro, A.~Ilic, N.~Roma, and P.~Tom{\'a}s, ``Modeling and decoupling the
  gpu power consumption for cross-domain dvfs,'' \emph{IEEE Transactions on
  Parallel and Distributed Systems}, 2019.

\bibitem{guerreiro2019dvfs}
J.~Guerreiro, A.~Ilic, N.~Roma, and P.~Tomás, ``Dvfs-aware application
  classification to improve gpgpus energy efficiency.'' \emph{Parallel
  Computing}, vol.~83, pp. 93--117, 2019.

\bibitem{lee2011heterogeneity}
G.~Lee and R.~H. Katz, ``Heterogeneity-aware resource allocation and scheduling
  in the cloud.'' in \emph{HotCloud}, 2011.

\bibitem{capodieci2018deadline}
N.~Capodieci, R.~Cavicchioli, M.~Bertogna, and A.~Paramakuru, ``Deadline-based
  scheduling for gpu with preemption support,'' in \emph{2018 IEEE Real-Time
  Systems Symposium (RTSS)}.\hskip 1em plus 0.5em minus 0.4em\relax IEEE, 2018,
  pp. 119--130.

\bibitem{Wang2019}
Q.~Wang and X.~Chu, ``{GPGPU Performance Estimation with Core and Memory
  Frequency Scaling},'' \emph{Proceedings of the International Conference on
  Parallel and Distributed Systems}, 2019.

\bibitem{Fan2019}
K.~Fan, B.~Juurlink, and T.~U. Berlin, ``{Predictable GPUs Frequency Scaling
  for Energy and Performance for Energy and Performance},'' no. Icpp, 2019.

\bibitem{Bridges2016}
R.~A. Bridges, N.~Imam, and T.~M. Mintz, ``{Understanding GPU power: A survey
  of profiling, modeling, and simulation methods},'' \emph{ACM Computing
  Surveys}, vol.~49, no.~3, 2016.

\bibitem{rodinia_che}
S.~{Che}, M.~{Boyer}, J.~{Meng}, D.~{Tarjan}, J.~W. {Sheaffer}, S.~{Lee}, and
  K.~{Skadron}, ``Rodinia: A benchmark suite for heterogeneous computing,'' in
  \emph{2009 IEEE International Symposium on Workload Characterization
  (IISWC)}, Oct 2009, pp. 44--54.

\bibitem{polybench_gray}
S.~Grauer-Gray, L.~Xu, R.~Searles, S.~Ayalasomayajula, and J.~Cavazos,
  ``Auto-tuning a high-level language targeted to gpu codes,'' in \emph{2012
  Innovative Parallel Computing (InPar)}.\hskip 1em plus 0.5em minus
  0.4em\relax IEEE, 2012, pp. 1--10.

\bibitem{basu2018interference}
A.~Basu, J.~L. Greathouse, G.~Venkataramani, and J.~Vesel{\`y}, ``Interference
  from gpu system service requests.'' in \emph{IISWC}, 2018, pp. 179--190.

\bibitem{allen2016variance}
Z.~Allen-Zhu and E.~Hazan, ``Variance reduction for faster non-convex
  optimization,'' in \emph{International conference on machine learning}, 2016.

\bibitem{Chau2017}
V.~Chau, X.~Chu, H.~Liu, and Y.~W. Leung, ``{Energy efficient job scheduling
  with DVFS for CPU-GPU heterogeneous systems},'' \emph{Proceedings of the 8th
  International Conference on Future Energy Systems}, 2017.

\bibitem{wu2015gpgpu}
G.~Wu, J.~L. Greathouse, A.~Lyashevsky, N.~Jayasena, and D.~Chiou, ``Gpgpu
  performance and power estimation using machine learning,'' in \emph{2015 IEEE
  21st International Symposium on High Performance Computer Architecture
  (HPCA)}.\hskip 1em plus 0.5em minus 0.4em\relax IEEE, 2015, pp. 564--576.

\bibitem{kriegel2009clustering}
H.-P. Kriegel, P.~Kr{\"o}ger, and A.~Zimek, ``Clustering high-dimensional data:
  A survey on subspace clustering, pattern-based clustering, and correlation
  clustering,'' \emph{ACM Transactions on Knowledge Discovery from Data
  (TKDD)}, vol.~3, no.~1, p.~1, 2009.

\bibitem{komoda2013power}
T.~Komoda, S.~Hayashi, T.~Nakada, S.~Miwa, and H.~Nakamura, ``Power capping of
  cpu-gpu heterogeneous systems through coordinating dvfs and task mapping,''
  in \emph{2013 IEEE 31st International Conference on Computer Design
  (ICCD)}.\hskip 1em plus 0.5em minus 0.4em\relax IEEE, 2013, pp. 349--356.

\bibitem{pardalos1987constrained}
P.~M. Pardalos and J.~B. Rosen, \emph{Constrained global optimization:
  algorithms and applications}.\hskip 1em plus 0.5em minus 0.4em\relax
  Springer, 1987, vol. 268.

\bibitem{Grid5000}
\BIBentryALTinterwordspacing
``Grid'5000,'' 2019. [Online]. Available:
  \url{https://www.grid5000.fr/w/Grid5000:Home}
\BIBentrySTDinterwordspacing

\bibitem{Basu2016}
P.~Basu, H.~Chen, S.~Saha, K.~Chakraborty, and S.~Roy, ``{SwiftGPU : Fostering
  Energy Efficiency in a Near-Threshold GPU Through a Tactical Performance
  Boost},'' \emph{Proceedings of the 2016 53rd ACM/EDAC/IEEE Design Automation
  Conference (DAC)}, 2016.

\bibitem{Teo2015}
Y.~M. Teo, ``{A model-driven approach for time-energy performance of parallel
  applications},'' \emph{2015 6th International Conference on Intelligent
  Systems, Modelling and Simulation}, pp. 3--4, 2015.

\bibitem{Ma2012}
K.~Ma, X.~Li, W.~Chen, C.~Zhang, and X.~Wang, ``{GreenGPU: A holistic approach
  to energy efficiency in GPU-CPU heterogeneous architectures},''
  \emph{Proceedings of the International Conference on Parallel Processing},
  pp. 48--57, 2012.

\bibitem{Losch2018}
A.~Losch and M.~Platzner, ``{A Highly Accurate Energy Model for Task Execution
  on Heterogeneous Compute Nodes},'' \emph{2018 IEEE 29th International
  Conference on Application-specific Systems, Architectures and Processors
  (ASAP)}, pp. 1--8, 2018.

\bibitem{Tang2019}
``{The impact of GPU DVFS on the energy and performance of deep Learning: An
  Empirical Study},'' \emph{e-Energy 2019 - Proceedings of the 10th ACM
  International Conference on Future Energy Systems}, no.~2, pp. 315--325,
  2019.

\end{thebibliography}

\newpage

\end{document}